\documentclass[aps,prd,reprint,nofootinbib,floatfix,superscriptaddress,secnumarabic,notitlepage]{revtex4-1}
\pdfoutput=1
\usepackage{setspace}
\usepackage{amsfonts,amssymb,epsfig,verbatim,mathbbol,amstext,amsmath,psfrag}
\usepackage{graphicx}
\usepackage[latin2]{inputenc}
\usepackage{t1enc}
\usepackage{amsmath}
\usepackage{subfigure}
\usepackage{dsfont}
\usepackage{slashed}
\usepackage{multirow}
\usepackage{wrapfig}
\usepackage{array}
\usepackage{enumerate}

\usepackage{colordvi}
\usepackage{rotfloat}
\usepackage{rotating}
\usepackage{color}
\usepackage[bottom]{footmisc}
\usepackage{MnSymbol}

\usepackage{cmap}

\usepackage[    bookmarks,
                 bookmarksopen = true,
                 bookmarksnumbered = true,
                 breaklinks = true,
                 linktocpage,
                 colorlinks = true,
                 linkcolor = blue,
                 urlcolor  = blue,
                 citecolor = blue,
                 anchorcolor = green,
                 hyperindex = true,
                 hyperfigures]
		{hyperref}  

\newcommand{\dd}{\textmd{d}}
\newcommand{\be}{\begin{equation}}
\newcommand{\ee}{\end{equation}}
\newcommand{\Tr}{\textmd{Tr}}
\newcommand{\Z}{\mathcal{Z}}

\newcommand{\D}{\mathcal{D}}

\renewcommand{\O}{\mathcal{O}}
\newcommand{\expv}[1]{\left \langle #1 \right \rangle}

\newcommand{\tr}{\textmd{tr}\,}

\newcommand{\Ml}{\mathcal{M}_{ud}}
\newcommand{\Ms}{\mathcal{M}_{s}}
\newcommand{\Dn}{\slashed{D}(0)}
\newcommand{\Dp}{\slashed{D}(\mu_I)}
\newcommand{\Dm}{\slashed{D}(-\mu_I)}
\newcommand{\ml}{m_{ud}}
\newcommand{\pion}{{\pi^\pm}}
\newcommand{\Nxi}{N}
\newcommand{\deltaNxi}{\delta^{\Nxi}}

\newcommand{\Tc}{T_{pc}}
\newcommand{\mut}{\mu_{I,pt}}
\newcommand{\Tt}{T_{pt}}

\long\def\symbolfootnote[#1]#2{\begingroup%
\def\thefootnote{\fnsymbol{footnote}}\footnote[#1]{#2}\endgroup} 

\newcommand{\Frankfurt}{Institute for Theoretical Physics, Goethe Universit\"at Frankfurt, D-60438 Frankfurt am Main, Germany}

\begin{document}
\title{QCD phase diagram for nonzero isospin-asymmetry}

\author{B.~B.~Brandt}
\affiliation{\Frankfurt}
\author{G.~Endr\H{o}di}
\affiliation{\Frankfurt}
\author{S.~Schmalzbauer}
\affiliation{\Frankfurt}

\begin{abstract}
The QCD phase diagram is studied in the presence of an isospin asymmetry using 
continuum extrapolated staggered quarks with physical masses. In particular, we investigate 
the phase boundary between the normal and the pion condensation phases and the 
chiral/deconfinement transition. The simulations are performed with a small explicit breaking 
parameter in order to avoid the accumulation of zero modes and thereby stabilize the 
algorithm. 
The limit of vanishing explicit breaking is obtained by means of an extrapolation, which 
is facilitated by a novel improvement program employing the singular value representation of the 
Dirac operator. 
Our findings indicate that no pion condensation takes place above $T\approx 160 \textmd{ MeV}$
and also suggest that the deconfinement crossover continuously connects to the BEC-BCS crossover at high 
isospin asymmetries.
The results may be directly compared to effective theories and model approaches to QCD.
\end{abstract}

\maketitle

\section{Introduction}

The thermodynamics of strongly interacting matter, as described by quantum chromodynamics (QCD),
plays a characteristic role for the phenomenology of heavy-ion 
collisions, the structure of compact stars and the evolution of the early Universe.
The relevant parameters that impact QCD physics in these settings   
include the temperature $T$ and the net densities $n_f$ of the individual quark flavors $f$. 
In the light quark sector ($f=u,d$), the characteristic combinations 
are the baryon density $n_B=(n_u+n_d)/3$ and the isospin density $n_I=n_u-n_d$. 
While the former measures the overall excess of strongly interacting matter over antimatter, 
the isospin density describes the asymmetry between the 
up and down quarks or, equivalently, between protons and neutrons in the system. 

The above-mentioned physical settings are all thought to exhibit a strong isospin asymmetry. 
The initial state of typical heavy-ion collisions has around twice as many neutrons as protons, which 
has implications, for example, for the imbalance between the generated charged pions~\cite{Li:1997px}. 
Cold neutron stars are characterized by even lower proton fractions~\cite{Steiner:2004fi}. 
A high isospin asymmetry is carried by charged pion degrees of freedom
that might be relevant for compact stars or for nuclear matter in general~\cite{Migdal:1990vm}.
Although the early Universe is typically assumed to undergo an evolution with almost vanishing
densities, a large lepton asymmetry might propagate into the baryon sector and shift equilibrium 
conditions around the time of the QCD transition~\cite{Schwarz:2009ii}. 

\begin{figure}[b]
 \centering
 \vspace*{-.5cm}
 \includegraphics[width=8.7cm]{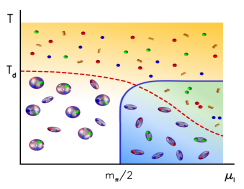}
 \caption{\label{fig:pd}A possible scenario for the 
 QCD phase diagram in the temperature -- isospin chemical potential
 plane. The different colors encode the various phases: hadronic phase (white),
 quark-gluon plasma (orange)
 pion condensation (blue) and BCS superconductivity (green). 
 The red dashed line indicates the deconfinement crossover and the blue
 solid line the second-order phase transition to the pion
 condensed phase.}
\end{figure}

In the grand canonical approach to QCD, the densities $n_f$ are traded for the corresponding chemical 
potentials $\mu_f$ and the isospin chemical potential is defined as $\mu_I=(\mu_u-\mu_d)/2$. 
The phase 
diagram in the temperature -- isospin chemical potential plane is known to exhibit at least four different 
phases that can be relevant for the physical systems mentioned above. 
We discuss these phases qualitatively in Fig.~\ref{fig:pd}. 
In the vacuum (for low values of $T$ and 
of $\mu_I$) QCD is confining and the effective degrees of freedom are hadrons. 
On the one hand, if the temperature is raised at low isospin chemical potential, 
deconfinement sets in and the quark-gluon plasma phase is realized. Lattice QCD simulations have 
demonstrated that the transition to deconfinement is 
an analytic crossover~\cite{Aoki:2006we,*Bhattacharya:2014ara} and occurs near the chiral pseudocritical temperature 
of $\Tc\approx 155 \textmd{ MeV}$~\cite{Borsanyi:2010bp,*Bazavov:2011nk}. 
On the other hand, if $\mu_I$ is increased at low temperature, another phase transition line is encountered,
which can be best understood using an effective description of QCD based on pionic 
degrees of freedom. 
Charged pions are the lightest hadrons that couple to the isospin chemical potential,
and their effective dynamics is described by chiral perturbation theory ($\chi$PT)~\cite{Son:2000xc}. 

At $T=0$, if the isospin chemical potential exceeds the critical value 
$\mu_{I,c}=m_\pi/2$, sufficient energy is pumped into the system so 
that charged pions can be created. Due to the bosonic nature of pions, a Bose-Einstein condensate (BEC) is 
formed at this point. $\chi$PT also predicts that the transition between the vacuum and 
the BEC state is of second order with the universality class $\mathrm{O}(2)$~\cite{Son:2000xc}.
As is usually the case, the temperature tends to destroy the condensate so that this phase transition line 
is expected to bend towards higher values of $\mu_I$ as $T$ grows. 
While $\chi$PT predicts a monotonous rise of $\mu_{I,c}(T)$, our preliminary lattice simulations 
reveal a flattening of this curve 
around the zero-density deconfinement crossover~\cite{Brandt:2016zdy,Brandt:2017zck}. This suggests 
that deconfinement is in some sense stronger than the condensation mechanism, and in the 
quark-gluon plasma phase no pions can be created by increasing $\mu_I$. 

Yet another transition is expected to occur at even higher isospin chemical potential. Perturbation theory 
predicts that the attractive gluon interaction forms Cooper-pairs (BCS superconductivity) 
of $u$ and $\bar d$ quarks in the 
pseudoscalar channel. Since the resulting pair has the same quantum numbers as the pion condensate, 
the deconfinement-type transition between the BEC and BCS states is expected to be an analytic crossover~\cite{Son:2000xc}.

The particular structure of the transition lines separating these 
four phases (hadronic, quark-gluon plasma, BEC and BCS)
in the phase diagram is not yet known. The results we present below favor the scenario 
depicted in Fig.~\ref{fig:pd},
where the $\mu_I=0$ deconfinement transition continuously connects to the BEC-BCS crossover
at high isospin chemical potentials.
We note that for asymptotically large values of $\mu_I$, perturbative arguments 
suggest~\cite{Son:2000xc,Cohen:2015soa} a decoupling of the gluonic sector and the emergence of 
a first-order deconfinement phase transition. This region is not included in Fig.~\ref{fig:pd}. 

The QCD phase diagram at nonzero temperature and isospin chemical potential 
has been studied using a multitude of 
approaches. Besides the already mentioned chiral perturbation theory as effective 
description~\cite{Son:2000xc,Splittorff:2000mm,*Loewe:2002tw,*Fraga:2008be,*Janssen:2015lda,*Carignano:2016lxe},
hard thermal loop perturbation theory~\cite{Andersen:2015eoa} 
and different low-energy models of QCD have been employed. 
The latter include the Nambu-Jona-Lasinio (NJL) 
model in four~\cite{Toublan:2003tt,*Frank:2003ve,*Barducci:2004tt,*He:2005sp,*Zhang:2006gu,*Sun:2007fc,*Xia:2013caa,*Zhang:2015baa,*Brauner:2016lkh,*Khunjua:2017mkc}
and in two dimensions~\cite{Gubina:2012wp}, the linear sigma or quark meson 
model~\cite{He:2005nk,*Ueda:2013sia,*Stiele:2013pma}, 
scalar field theory~\cite{Andersen:2006ys}, random matrix models~\cite{Klein:2003fy}, 
the thermodynamic bag model~\cite{Klahn:2016uce}, 
the hadron resonance gas model~\cite{Toublan:2004ks}
and by means of the functional renormalization group or Dyson-Schwinger
equations~\cite{Zhang:2006dn,*Svanes:2010we,*Kamikado:2012bt,*Wang:2017vis}. 
The impact of the isospin chemical potential 
was also investigated in the limit of large number $N_c$ of 
colors~\cite{Kashiwa:2017yvy} and employing the holographic principle~\cite{Aharony:2007uu,*Basu:2008bh,*Albrecht:2010eg}.
For large $N_c$, the phase diagram was argued to be related to the analogous phase diagram 
with baryon chemical potential via the orbifold equivalence~\cite{Hanada:2011ju}. 
In particular, this equivalence is expected to hold outside the pion condensation phase only, so that 
the precise knowledge of the location of the BEC transition line is relevant from this point of view 
as well.

A significant advantage of the $\mu_I>0$ system is that 
-- unlike for nonzero baryon chemical potential -- there is no sign problem and 
lattice QCD simulations can be employed to investigate the phase diagram directly. 
Following the pioneering work of 
Refs.~\cite{Kogut:2002tm,Kogut:2002zg}, 
the phase diagram was studied both in 
the grand canonical~\cite{Kogut:2004zg,Cea:2012ev,Endrodi:2014lja} and in the 
canonical ensemble~\cite{deForcrand:2007uz,Detmold:2012wc}. 
The low isospin density region was also discussed by means of a Taylor expansion in
$\mu_I$~\cite{Allton:2005gk,Borsanyi:2011sw}, while the strong coupling regime 
was investigated using the strong coupling expansion~\cite{Nishida:2003fb,*Langelage:2014vpa}.
Finally, we mention that QCD with $\mu_I>0$ has deep relations 
to two-color QCD with baryon chemical potentials. The latter theory has also been the subject of 
intensive research, both on the lattice~\cite{Kogut:2001na,*Hands:2006ve,*Bornyakov:2017txe} as well as in 
model and functional 
approaches~\cite{Andersen:2010vu,*Strodthoff:2011tz,*vonSmekal:2012vx}. 

We mention that, besides the broad range of applications for this system, 
an additional motivation to study the $\mu_I>0$ theory is its similarity 
with nonvanishing baryon chemical potential, $\mu_B>0$, both on the conceptual and on the technical level. 
At $T=0$, both systems exhibit the so-called silver blaze phenomenon~\cite{Cohen:2003kd} and 
undergo a phase transition from the vacuum to a phase where colorless composite particles are created.
This transition is accompanied by the accumulation of near-zero 
eigenvalues of the fermion matrix in both cases, leading to an ill-conditioned inversion 
problem and a breakdown of naive simulation algorithms. 
This necessitates the use of an infrared regulator that we denote by $\lambda$
below. 
Understanding these concepts and facing these technical challenges 
in the (sign-problem-free) $\mu_I\neq0$ theory may give us insight on how to
assess 
the $\mu_B\neq0$ system in the future. 

In this paper we follow the grand canonical approach and simulate QCD with $2+1$ flavors 
of dynamical staggered quarks at various values of $T$ and $\mu_I$. 
Besides the inclusion of the strange quark, we improve 
over currently existing studies in the literature by 
using physical quark masses and employing an improved lattice action in the simulations.
The latter enables a faster scaling towards the continuum limit. 
Furthermore, we present a novel method to perform the extrapolation of the infrared 
regulator $\lambda$ to zero -- the main technical achievement in this project. 
This 
method involves the singular values of the massive Dirac operator, which 
we discuss for the first time on the lattice in this context.
Preliminary results of our simulations were presented in Refs.~\cite{Brandt:2016zdy,Brandt:2017zck} and applied
in Ref.~\cite{Bali:2016nqn}.

\section{Setup and observables}

\subsection{Symmetries}
\label{sec:sb}

First of all, it is instructive to discuss the pattern of the symmetry breaking that 
leads to Bose-Einstein condensation at high isospin chemical potentials. 
In Euclidean spacetime, our choice for the continuum action $S_{ud}=\bar\psi\Ml\,\psi$ 
for the light quarks $\psi=(u,d)^\top$ includes
\be
\Ml = \gamma_\mu (\partial_\mu + i A_\mu)\, \mathds{1} + \ml \mathds{1} + \mu_I \gamma_4 \tau_3  + i \lambda \gamma_5 \tau_2\,.
\label{eq:Sud}
\ee
Here $A_\mu$ is the gluon field and $\tau_a$ denote the Pauli matrices. Besides the isospin chemical potential $\mu_I$
and the quark mass $m_{ud}$,  
we also included an explicit symmetry breaking term in $\Ml$ that couples to 
the charged pion field $\pion$,
\be
S_{ud}=S_{ud}(\lambda=0)+\lambda\,\pion, \quad
\pion \equiv \bar\psi i\gamma_5\tau_2 \psi = \bar u \gamma_5 d - \bar d \gamma_5 u\,.
\ee
The role of the parameter $\lambda$ -- referred to as a pionic source in the following -- 
will be elucidated below. 

At $\mu_I=\lambda=0$ (but nonzero $\ml)$, the action $S_{ud}$ is symmetric under 
the chiral group 
$\mathrm{SU}_V(2)\times \mathrm{U}_V(1)$.
The $\mathrm{U}_V(1)$ symmetry, corresponding to baryon number conservation, is not affected by the 
chemical potential nor by the pionic source and is not discussed in the following. 
The $\mathrm{SU}_V(2)$ symmetry group is broken down to
$\mathrm{U}_{\tau_3}\!(1)$ by the chemical potential. 
The subscript indicates that the generator of the remaining
symmetry 
is $\tau_3$. 
Pion condensation is signaled by the spontaneous breaking of this 
$\mathrm{U}_{\tau_3}\!(1)$ symmetry, by any of the expectation
values 
$\expv{\bar\psi \gamma_5 \tau_1 \psi}$, $\expv{\bar\psi \gamma_5 \tau_2 \psi}$.

The {\it spontaneous} 
breaking of the continuous $\mathrm{U}_{\tau_3}\!(1)$ symmetry implies the presence of a 
Goldstone mode. The introduction of the pionic source $\lambda$ in~(\ref{eq:Sud}) corresponds to an additional {\it explicit} breaking that
selects the $\tau_2$ direction for the ground state and makes the would-be massless mode
a pseudo-Goldstone boson. In fact, in a finite volume, such a trigger is necessary 
for the spontaneous breaking to occur. 
The physical limit $\lambda\to0$ is obtained subsequently by means of an extrapolation.
Notice that for the $\mathrm{U}_{\tau_3}\!(1)$ symmetry, its spontaneous (by $\expv{\pion}$) and 
explicit (by $\lambda$) breaking is completely analogous to the one of the standard 
chiral symmetry at $\mu_I=0$, together with its spontaneous (by $\expv{\bar\psi\psi}$)
and explicit (by $\ml$) breakings. One very important difference is that while in nature 
$\ml>0$, the parameter $\lambda$ is unphysical and the limit $\lambda\to0$ needs to be taken.

For vanishing isospin chemical potential, the pionic source can be rotated into the mass 
parameter. In particular, we can perform 
the chiral rotation
\be
\psi\to e^{i\alpha \gamma_5 \tau_2/2 } \psi, \quad \bar\psi \to \bar\psi\, e^{i\alpha \gamma_5 \tau_2/2 },\quad
\alpha=\arctan \frac{\lambda}{\ml}\,.
\ee
This rotation brings the action into a flavor-diagonal form
\be
S_{ud} = \bar \psi \Ml \psi, \quad\quad \Ml = \gamma_\mu (\partial_\mu + i A_\mu)\;\mathds{1} + \sqrt{\ml^2+\lambda^2} \;\mathds{1}\,.
\label{eq:zeromui}
\ee
Thus, at $\mu_I=0$ the mass and the pionic source parameter are indistinguishable
and only their squared sum plays a physical role. 
This will have important consequences for the ultraviolet structure of the theory, 
see Sec.~\ref{sec:obsrenorm} below. 
Notice that for $\mu_I=0$ the pionic source has the same interpretation as 
the twisted mass parameter used in the context of Wilson fermions~\cite{Frezzotti:2000nk}. 

\subsection{Lattice setup}
\label{sec:setup} 
 
To investigate pion condensation and the associated spontaneous symmetry breaking, 
we study 2+1-flavor QCD with $\mu_I>0$ and $\lambda>0$ nonperturbatively on the lattice.
We work on $N_s^3\times N_t$ lattices with spacing $a$ and sites indexed 
by the coordinates $(n_x,n_y,n_z,n_t)$. 
The temperature and the spatial volume read $T=1/(N_ta)$ and $V=(N_sa)^3$, respectively. 
To discretize the Dirac operator we use 
the staggered formulation; here the equivalent of $\gamma_5$ is the local 
spin-flavor structure $\eta_5 = \gamma_5^{\rm S} \otimes \gamma_5^{\rm F}=(-1)^{n_x+n_y+n_z+n_t}$. At $\mu_I=\lambda=0$, 
this operator couples to the Goldstone boson in the chiral limit. 

The partition function of this system is given in terms of the path integral over
the gluon links $U_\mu=\exp(iaA_\mu)$,
\be
\Z = \int \D U_\mu \, e^{-\beta S_G}\, (\det \Ml)^{1/4}\,(\det \Ms)^{1/4} \,,
\label{eq:Z}
\ee
where we employed the rooting procedure (for a discussion on the theoretical issues 
related to rooting, see Ref.~\cite{Durr:2005ax}). In Eq.~(\ref{eq:Z}), $\beta=6/g^2$ denotes the inverse gauge coupling, 
$S_G$ is the tree-level Symanzik improved gluon action, $\Ml$ is the light quark matrix in the basis
of the up and down quarks and $\Ms$ is the strange quark matrix,
\be
\Ml
\!= \!
\begin{pmatrix}
 \Dp+\ml & \!\!\!\lambda \eta_5 \\
 -\lambda \eta_5 & \!\!\!\Dm +\ml
\end{pmatrix}, \;\;\,
\Ms \!=\! \Dn + m_s\,.
\label{eq:M}
\ee
Here, the argument of $\slashed{D}$ indicates the chemical potential $\mu_I$, introduced on the 
lattice by multiplying the forward (backward) timelike links by $e^{a\mu_I}$ ($e^{-a\mu_I}$). 
The pionic source parameter $\lambda$ induces an explicit symmetry breaking and serves 
to trigger pion condensation as we discussed above in Sec.~\ref{sec:sb}.

The integrand of $\Z$ needs to be positive to allow the importance sampling of the 
gluon configurations. 
The strange quark determinant is real and positive due to the standard $\eta_5$-hermiticity
relation $\eta_5 \Ms \eta_5 = \Ms^\dagger$. To show the positivity for the light sector, we need 
to discuss the symmetry properties of the staggered Dirac operator at $\mu_I\neq0$. 
The staggered equivalent of chiral symmetry implies that
\be
\Dp \eta_5 + \eta_5 \Dp = 0\,
\label{eq:chs}
\ee
holds. In addition, 
the Dirac operator satisfies
\be
\eta_5 \Dp\eta_5 = \Dm^\dagger \,.
\label{eq:g5h}
\ee
so that the light fermion matrix is $\tau_1 \eta_5$-hermitian
\be
\tau_1\eta_5 \,\Ml \,\eta_5 \tau_1 = \Ml^\dagger\,.
\ee
Thus, taking the determinant of both sides shows that $\det \Ml$ is real.

One can also show that the determinant is positive by 
considering
\be
\Ml'=B\Ml B = 
\begin{pmatrix}
 \Dp+\ml & \lambda \\
 -\lambda & [\Dp+\ml]^\dagger \\
\end{pmatrix},
\label{eq:defB}
\ee
where $B=\textmd{diag}(1,\eta_5)$ and we used Eq.~(\ref{eq:g5h}).
Indeed, since $B$ has unit determinant, we have
\be
\det \Ml = \det\left( [\Dp+\ml][\Dp+\ml]^\dagger+\lambda^2\right) >0\,.
\label{eq:detMM}
\ee
Thus, both determinants in the measure of the path integral~(\ref{eq:Z}) are positive. 

The fourth root of the determinants in~(\ref{eq:Z})
is approximated via rational functions. The simulation setup with $\lambda>0$ was first 
introduced for the $N_f=2$ theory in the 
pioneering work of Ref.~\cite{Kogut:2002tm} for the quenched case and in 
Ref.~\cite{Kogut:2002zg} for dynamical QCD. The same technique was also used in Ref.~\cite{Endrodi:2014lja}.
Here we extend this setup by including the strange quark as well. In addition, 
we improve the lattice action by using the tree-level Symanzik gauge action and by 
employing two steps of stout smearing in the Dirac operator. 
The quark masses are tuned to their physical values along the line 
of constant physics (LCP) $m_f(\beta)$, as determined in Ref.~\cite{Borsanyi:2010cj}, 
with the pion mass $m_\pi\approx 135\textmd{ MeV}$.
Our simulation algorithm is based on Ref.~\cite{Aoki:2005vt}. In addition we implement 
a Hasenbusch-type improvement scheme that is typically used in the context of mass 
preconditioning~\cite{Urbach:2005ji}. In our setup this amounts to the replacement 
$\det \Ml(\lambda) = \det \Ml(\rho)\, \cdot\, \det \Ml(\lambda)/\det \Ml(\rho)$ with $\rho>\lambda$, 
which allows to use a larger step size (and a lower precision) in 
the simulation algorithm for the second (more expensive) factor.

\subsection{Observables and renormalization}
\label{sec:obsrenorm}

Our primary observables are the pion condensate and the quark condensate. 
Both are obtained from the partition function via differentiation,
\be
\expv{\pion}
= \frac{T}{V}\frac{\partial \log\Z}{\partial \lambda}, \quad\quad\quad
\expv{\bar\psi\psi} = \frac{T}{V}\frac{\partial \log\Z}{\partial \ml}\,.
\label{eq:pbpdef}
\ee
Inserting the partition function~(\ref{eq:Z}) and 
rewriting the light quark determinant using Eq.~(\ref{eq:detMM}), we obtain
for the condensate operators,
\be
\begin{split}
\pion &= \frac{T}{2V} \,\tr
\frac{\lambda}{|\Dp+\ml|^2+\lambda^2}, \\
\bar\psi\psi &= \frac{T}{2V} \,\textmd{Re } \tr
\frac{\Dp+\ml}{|\Dp+\ml|^2+\lambda^2}\,.
\end{split}
\label{eq:traces1a}
\ee
The relation~(\ref{eq:traces1a})
allows for direct measurements using noisy estimators. 

The observables of Eq.~(\ref{eq:pbpdef}) are subject to additive renormalization.
This is necessary, since $\log\Z$ contains ultraviolet divergences in the 
inverse lattice spacing. The structure of these divergences 
can be determined based on dimensional arguments
(see Ref.~\cite{Leutwyler:1992yt} for a discussion at $\mu_I=\lambda=0$),
\be
\log\Z \sim a^{-4} + (\ml^2+\lambda^2) \,a^{-2} + (\ml^2+\lambda^2)^2 \log a\,,
\label{eq:divs}
\ee
where we suppressed further divergences that contain $m_s$. Note that the divergences 
are independent of $\mu_I$, since the chemical potential couples to a conserved 
charge~\cite{Hasenfratz:1983ba}. Thus, it suffices to consider the case of $\mu_I=0$.
Above in Eq.~(\ref{eq:zeromui}) we have seen that for vanishing isospin chemical potential, the mass and the pionic source may be rotated into each other, so that the 
two parameters can only appear in Eq.~(\ref{eq:divs}) in the form $\ml^2+\lambda^2$. 
The quark condensate and the pion condensate inherit the quadratic and the 
logarithmic divergences from $\log\Z$. However, for $\expv{\pion}$ these vanish at $\lambda=0$ -- the point 
of interest for the physical theory. Since the light quark mass
is nonzero, the additive divergences remain in $\expv{\bar\psi\psi}$ and need to be subtracted even in the 
limit $\lambda\to0$. The standard 
choice is to consider the difference to $\expv{\bar\psi\psi}$ measured in the vacuum, 
i.e.\ at $T=\mu_I=0$. 

Finally we need to address the multiplicative renormalization of our observables.
The quark condensate and the pion condensate have nontrivial renormalization constants, 
$Z_{\bar\psi\psi}= Z_{\ml}^{-1}$ and $Z_\pi = Z_\lambda^{-1}$. 
The equivalence of $\ml$ and $\lambda$ at $\mu_I=0$ and the $\mu_I$-independence of 
the renormalization constants, however, imply that $Z_{\ml}=Z_{\lambda}$, so that 
a multiplication of the condensates by $\ml$ cancels the multiplicative
divergence.\footnote{Note that multiplying by $\lambda$ would also result in a renormalization 
group invariant combination. However, this construction vanishes in the $\lambda\to0$ limit and 
is therefore not useful.}
Altogether, the renormalized observables read
\be
\begin{split}
\Sigma_{\bar\psi\psi} &= \frac{\ml}{m_\pi^2 f_\pi^2} \left[ \expv{\bar\psi\psi}_{T,\mu_I} - \expv{\bar\psi\psi}_{0,0} \right] + 1, \\ 
\Sigma_{\pi} &= \frac{\ml}{m_\pi^2 f_\pi^2} \expv{\pion}_{T,\mu_I}\,,
\end{split}
\label{eq:renobs}
\ee
where we also included a normalization factor involving the pion 
mass $m_\pi=135 \textmd{ MeV}$ and 
the chiral limit of the pion decay constant
$f_\pi=86 \textmd{ MeV}$ and added unity to the quark condensate 
for convenience. In this normalization (which follows Ref.~\cite{Bali:2012zg}), 
$\Sigma_{\bar\psi\psi}=1$ at $T=\mu_I=0$ due to the Gell-Mann-Oakes-Renner relation. In addition, 
zero-temperature 
leading-order $\chi$PT~\cite{Son:2000xc} predicts a gradual rotation of the condensates so that 
$\Sigma_{\bar\psi\psi}^2 + \Sigma_\pi^2=1$ holds irrespective of $\mu_I$,
which can also be observed to some extent in the full theory. 

In addition to the fermionic observables we also consider the Polyakov loop,
\be
P = \expv{ \frac{1}{V}\! \sum\limits_{n_x,n_y,n_z}\! \Tr \prod\limits_{n_t=0}^{N_t-1} U_t(n) }\,,
\ee
as a measure for deconfinement. 
The multiplicative renormalization of $P$ amounts to
\be
P_r (T,\mu_I) 
=  Z
\cdot P(T,\mu_I),
\quad\;
Z
= \left(\frac{P_\star}{P(T_\star,\mu_I=0)} \right)^{T_\star / T},
\label{eq:pren}
\ee
with an arbitrary choice for $P_\star=P_r(T_\star,0)$ and $T_\star$. Different choices 
correspond to different renormalization schemes; we choose 
$T_\star = 162 \textmd{ MeV}$, where the bare Polyakov loops are already significantly nonzero, and set $P_\star=1$.
This renormalization prescription for $P$ was developed
at $\mu_I=0$~\cite{Borsanyi:2012uq} and also put into practice 
at nonzero background magnetic fields~\cite{Bruckmann:2013oba}.

\section{Improvements for the \boldmath$\lambda\to0$ extrapolation}
\label{sec:imprl}

\begin{figure}[t]
 \centering
 \includegraphics[width=8cm]{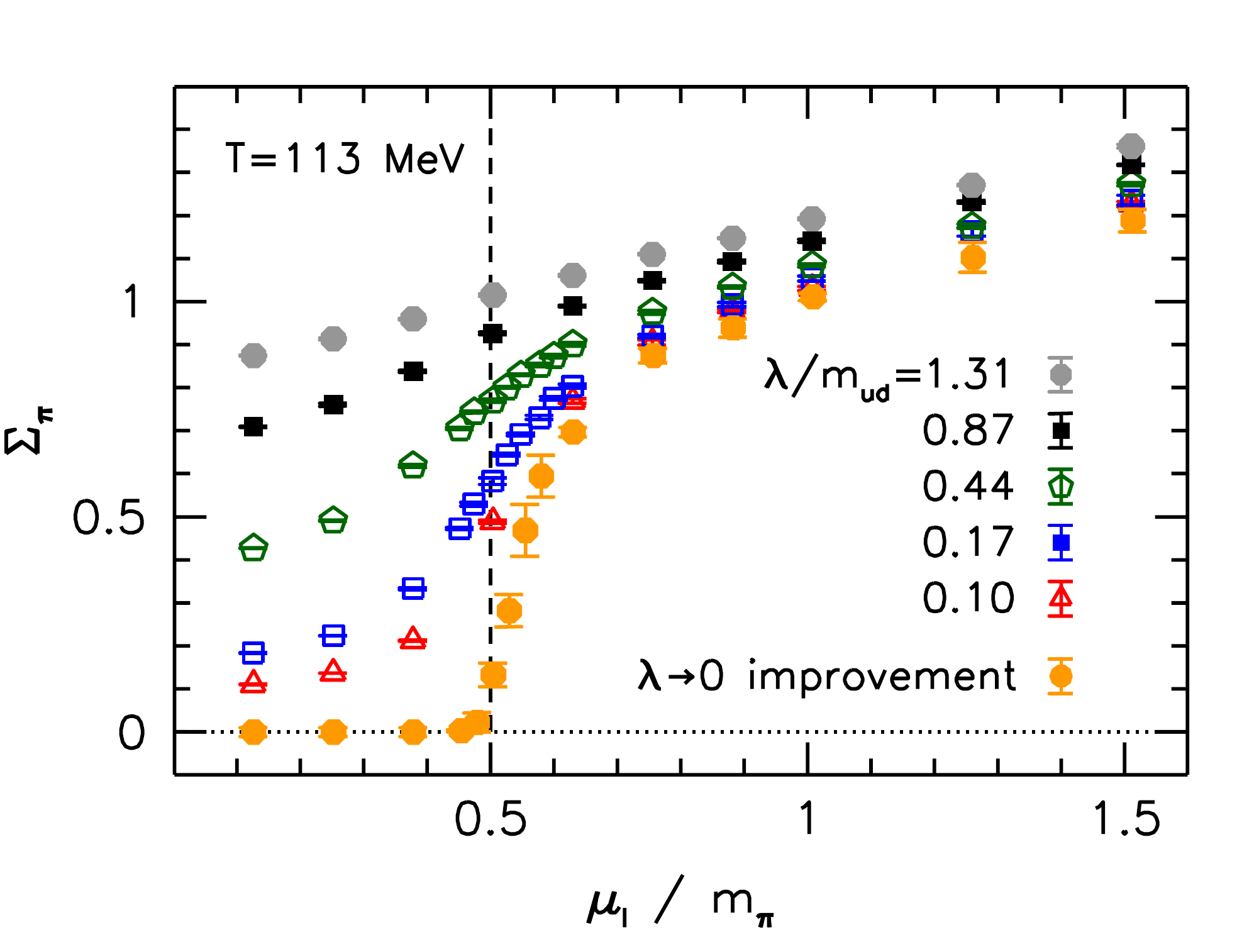}\quad
 \includegraphics[width=8cm]{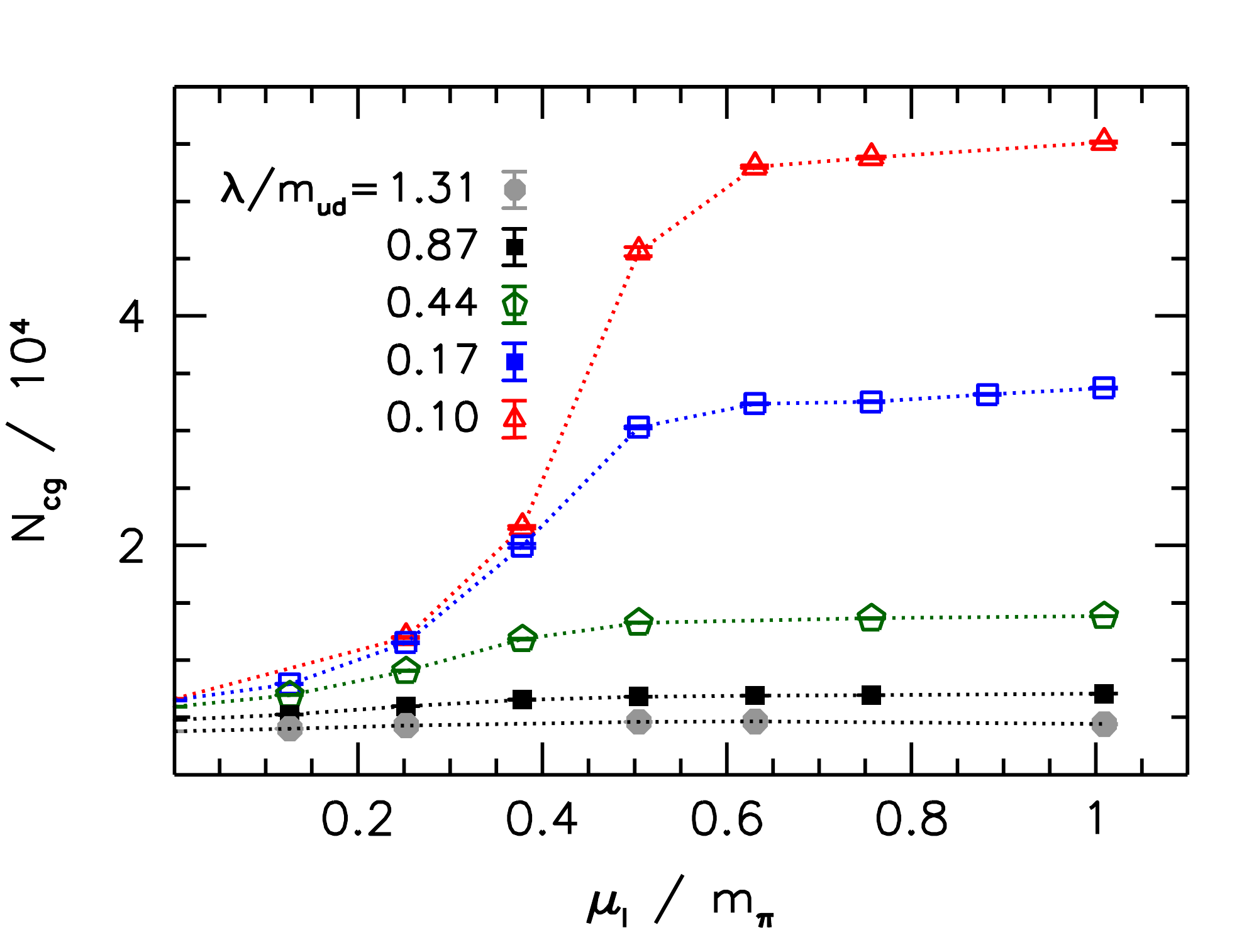}
 \caption{\label{fig:pbGp_impro}
 Top panel: direct results for the pion condensate using the representation~(\ref{eq:traces1a}), 
 obtained on ensembles with 
 different values of the pionic source $\lambda$ (gray to red points).
 For comparison, the results of our improved $\lambda\to0$ extrapolation are also shown (yellow points).
 Bottom panel: average iteration count for inverting $\Ml$ for various values of the isospin chemical potential
 and the pionic source. 
 }
\end{figure}

The pionic source parameter $\lambda>0$ is introduced in order to trigger pion condensation 
and to stabilize the simulation algorithm. However, in order to obtain physical results 
$\lambda$ needs to be extrapolated to zero. Most of the 
observables exhibit a pronounced dependence on $\lambda$, making this extrapolation cumbersome. 
As a typical example, in the top panel of Fig.~\ref{fig:pbGp_impro} we show the pion condensate $\Sigma_{\pi}$
of Eq.~(\ref{eq:renobs}) as measured using various values of $\lambda$ on our $24^3\times 6$ 
ensembles at $T=113 \textmd{ MeV}$. Especially below and around the critical chemical potential
$\mu_{I,c}\approx m_\pi/2$, a direct fitting of the data appears to be hopeless.
This necessitates various 
improvements, both in the valence sector (the definition of the operators) and 
in the sea sector (reweighting of the configurations), to eliminate the $\lambda$-dependence 
in the observables. The result of the improvement, which we will describe below, is also included in the top panel of Fig.~\ref{fig:pbGp_impro} -- clearly 
revealing the transition between the vacuum and the BEC phase, which was hidden for the direct data at high $\lambda$.

The simulations at low values of $\lambda$ suffer from a numerical problem. 
Since the pionic source acts as an infrared regulator, it has a substantial impact on the condition number of 
the fermion matrix $\Ml$. In particular, the average iteration count $N_{\rm cg}$ for the convergence of the conjugate gradient
algorithm used for inverting $\Ml$ grows significantly as $\lambda$ decreases. This is especially the case in the 
pion condensed phase (see the bottom panel of Fig.~\ref{fig:pbGp_impro}), where the inversion at our smallest $\lambda$
is an order of magnitude slower than at the largest pionic source.\footnote{We also mention that simulating with $\lambda=0$ is not feasible even for our smallest isospin chemical potentials
due to the occasional appearance of configurations with fermion matrices of 
very high condition numbers, for which the numerical inversion breaks down.} 
Moreover, in order to maintain a reasonable acceptance in the Metropolis step, 
the simulations at low $\lambda$ need to be performed with a reduced step size 
along the Monte-Carlo trajectory due to increasing fluctuations in the fermion force. This leads to a
slowing down by a further factor of $3-4$ for the lowest pionic source.

Thus, improving our observables so that they lie closer to their $\lambda\to0$ limit 
is necessary -- both for controlling the extrapolation and for sparing simulation time. 
It turns out that the improvement
for the pion condensate is drastically different from the improvement of the other 
observables due to the fact that $\expv{\pion}$ plays the role of the order parameter 
for Bose-Einstein condensation and thus has a highly nontrivial finite volume scaling. 
There is, nevertheless, a common element in the improvement program, which is the singular value representation of the massive Dirac operator.

\subsection{Singular values}

To get acquainted with the notion of singular values, let us 
begin with the eigenvalue equation of the massive Dirac operator. For the 
up quark, the eigensystem reads
\be
[\Dp+\ml] \,\psi_n = (\nu_n+\ml)\,\psi_n\,,
\label{eq:Dpp}
\ee
where the eigenvalues $\nu_n$ are complex numbers. Using 
chiral symmetry~(\ref{eq:chs}) and the hermiticity relation~(\ref{eq:g5h}) we obtain 
the eigensystem for the down quark,
\be
{\widetilde\psi_n}^\dagger \,[\Dm+\ml] = {\widetilde\psi_n}^\dagger(\nu_n^*+\ml)\,,\quad\quad
\widetilde\psi_n = \eta_5\psi_n\,.
\label{eq:Dmm}
\ee
Note
that $\Dp$ is not a normal operator thus $[\Dp,\slashed{D}^\dagger(\mu_I) ] \neq0$ and its left 
and right eigenvectors do not coincide. Nevertheless, Eqs.~(\ref{eq:Dpp}) and~(\ref{eq:Dmm}) reveal 
that for each eigenvalue in the up quark sector there is a complex conjugate pair in the 
down quark sector, which is why the determinant of the total light quark matrix is real and 
positive as we have seen in Sec.~\ref{sec:setup}.

We can create a hermitian operator by taking the modulus squared of the Dirac operator~\cite{Kanazawa:2011tt}. 
The square roots of the eigenvalues of this operator are referred to as the {\it singular values}\footnote{The name refers to the role of these in the singular 
value decomposition of non-hermitian matrices.} of the 
Dirac operator. 
The eigenvalue equation reads
\be
[\Dp+\ml]^\dagger[\Dp+\ml]\, \varphi_n = \xi_n^2 \,\varphi_n\,.
\label{eq:sval}
\ee
Note that due to the non-normality of $\Dp$, there is no relation 
between the singular values $\xi_n$ and the 
eigenvalues $\nu_n$.
We note that the singular values of $\Dp+\ml$ and of $\Dm+\ml$ coincide due to
the hermiticity relation~(\ref{eq:g5h}).

\subsection{Banks-Casher type relation for the pion condensate}
\label{sec:BCpi}

An important characteristic of chiral symmetry breaking at $\mu_I=0$ is reflected 
by the Banks-Casher relation~\cite{Banks:1979yr}, which states that 
the chiral limit of the quark condensate is related to the 
density of the Dirac eigenvalues around zero. Below we demonstrate that 
the $\lambda\to0$ limit of the pion condensate can be similarly obtained by looking 
at the density of the singular values around zero. 
The derivation follows Ref.~\cite{Kanazawa:2011tt}, where 
massless quarks at $\mu_I>0$ were considered. Here we generalize the discussion 
to arbitrary $\ml$. Throughout this section we neglect exact zero modes, whose 
contribution is discussed in detail in Ref.~\cite{Kanazawa:2011tt}.

Using Eq.~(\ref{eq:sval}), we can write down the singular value representation of the 
pion condensate. Writing the trace of Eq.~(\ref{eq:traces1a}) in the basis 
of the $\varphi_n$ modes we obtain,
\be
\begin{split}
\expv{\pion} &= \frac{\lambda T}{2V} \expv{\sum_n \,(\xi_n^2+\lambda^2)^{-1}}\\
&\xrightarrow{V\to\infty} \frac{\lambda }{2} \expv{\int \dd \xi \,\rho(\xi)
(\xi^2+\lambda^2)^{-1}} \\
&\xrightarrow{\lambda\to0} \frac{\pi }{4} \expv{\rho(0)}\,.
\end{split}
\label{eq:BC}
\ee
Here in the first step we considered the volume to be large enough so that 
the singular values become sufficiently dense and the sum can be replaced by 
an integral introducing the density of the singular values,
\be
\expv{\rho(\xi)} = \lim_{V\to\infty} \frac{T}{V} \expv{\sum_n \delta(\xi-\xi_n)}\,.
\ee
In the second step of Eq.~(\ref{eq:BC}) we performed the $\lambda\to0$ limit, which lead to a 
representation of the $\delta$-function and resulted in the density $\rho(0)$
around zero.
Eq.~(\ref{eq:BC}) tells us that having a nonzero pion condensate is equivalent to an 
accumulation of the near-zero singular values of the massive Dirac operator.
Notice that the ordering of the two limits in Eq.~(\ref{eq:BC}) -- just like in the 
usual Banks-Casher relation -- is essential, and setting $\lambda=0$ in the operator 
directly would give $\pion=0$. 
Equation~(\ref{eq:BC}) also connects the 
slowing down of the operator inversion and the high condition number of the fermion matrix
in the BEC phase (as 
we demonstrated in the bottom panel of Fig.~\ref{fig:pbGp_impro})
with a physical notion: the emergence of a nonzero pion condensate. 

\subsection{Improvement of the quark condensate}
\label{sec:obsimpr}

For the quark condensate operator
of Eq.~(\ref{eq:traces1a}),
the spectral representations reads
\be
\bar\psi\psi = \frac{T}{2V} \sum_n \frac{\textmd{Re}\; \varphi_n^\dagger [\Dp+\ml] \varphi_n}{\xi^2_n+\lambda^2}\,.
\ee
In this case the $\lambda\to0$ limit can be taken explicitly, without approaching the 
thermodynamic limit first. However, unlike the pion condensate, the quark condensate necessitates the calculation of 
various matrix elements and not only that of the singular values. 
Besides the values of the operators at $\lambda=0$, it 
will be advantageous to have access to the $\lambda$ dependence of the operator as well.
In particular, we define the change in the operator between $\lambda$ and $\lambda=0$,
\be
\begin{split}
\delta_{\bar\psi\psi} &\equiv \bar\psi\psi(\lambda)-\bar\psi\psi(\lambda=0) \\
&= 
\frac{T}{2V} \sum_n \textmd{Re}\; \varphi_n^\dagger [\Dp+\ml] \varphi_n \cdot
\left( \frac{1}{\xi_n^2+\lambda^2}-\frac{1}{\xi_n^2}\right)\,.
\end{split}
\label{eq:ldiffs}
\ee
We will use $\delta_{\bar\psi\psi}$ to improve the $\lambda$ dependence 
of the $\bar\psi\psi$ operator calculated using noisy estimators. 

\subsection{Leading-order reweighting}
\label{sec:reweight}

The above improvements eliminated the explicit dependence of the operators (i.e., of the valence 
quarks) on the 
pionic source. The remaining $\lambda$-dependence originates from sea 
quarks, i.e.\ from the nonzero value of $\lambda$ in the path
integral measure used to define expectation values $\expv{\O}_{\lambda}$.
In particular, this involves the $\lambda$-dependence of the light quark determinant in Eq.~(\ref{eq:Z}). 
To get rid of this contribution, we need to manipulate the distribution of configurations
by introducing the reweighting factors,
\be
\begin{split}
\expv{\O}_{\lambda=0} &= \frac{\expv{\O \,W(\lambda)}_{\lambda>0} }{ \expv{W(\lambda)}_{\lambda>0} },\\
W(\lambda) &\equiv \frac{\det \left[|\Dp+\ml|^2\right]^{1/4}}{\det\left[ |\Dp+\ml|^2+\lambda^2\right] ^{1/4}}\,,
\end{split}
\label{eq:rewe}
\ee
where we used Eq.~(\ref{eq:detMM}). This way the $\lambda>0$ determinant is
canceled in the expectation values so that we 
mimic the distribution that would have been obtained via a simulation directly at $\lambda=0$.
Note that Eq.~(\ref{eq:rewe}) only holds if the distributions 
at $\lambda>0$ and at $\lambda=0$ have sufficiently large overlap. 

The full determinant is a very expensive object as in principle 
it necessitates the calculation of all 
singular values. However, since we only need $W(\lambda)$ for small pionic sources, 
we can expand the reweighting factor in $\lambda$. 
Rewriting the logarithm of the reweighting factor in this manner, we obtain
\begin{align}
\log W(\lambda) &= \log \left.\frac{\det
\left[|\Dp+\ml|^2+\lambda^2-\lambda_w^2\right]^{1/4}}{\det\left[
|\Dp+\ml|^2+\lambda^2\right] ^{1/4}}\right|_{\lambda_w=\lambda}  \nonumber \\
&= \left[ -\frac{\lambda_w^2}{4} \;\tr \frac{1}{|\Dp+\ml|^2+\lambda^2} + \O(\lambda_w^4) \right]_{\lambda_w=\lambda}
\nonumber \\
&=
\left[ -\lambda_w^2\frac{V}{2T} \frac{\pion}{\lambda} + \O(\lambda_w^4)
\right]_{\lambda_w=\lambda} \nonumber \\
&= -\frac{\lambda V}{2T}\pion + \O(\lambda^4) \equiv \log W_{\rm LO}(\lambda) + \O(\lambda^4)\,.
\label{eq:LOrewe}
\end{align}
More specifically, here we replaced the logarithm of the determinant ratio 
with its Taylor-expansion in 
the pionic source (the expansion variable is denoted $\lambda_w$)
and evaluated the result at $\lambda_w=\lambda$.
We then exploited the fact that 
odd terms in the expansion vanish. 
On comparison with Eq.~(\ref{eq:traces1a}), 
the leading order term in the expansion is found to be proportional to the 
pion condensate. The resulting leading-order reweighting factor is denoted by $W_{\rm LO}(\lambda)$. 

\begin{figure}[ht!]
 \centering
 \includegraphics[width=8cm]{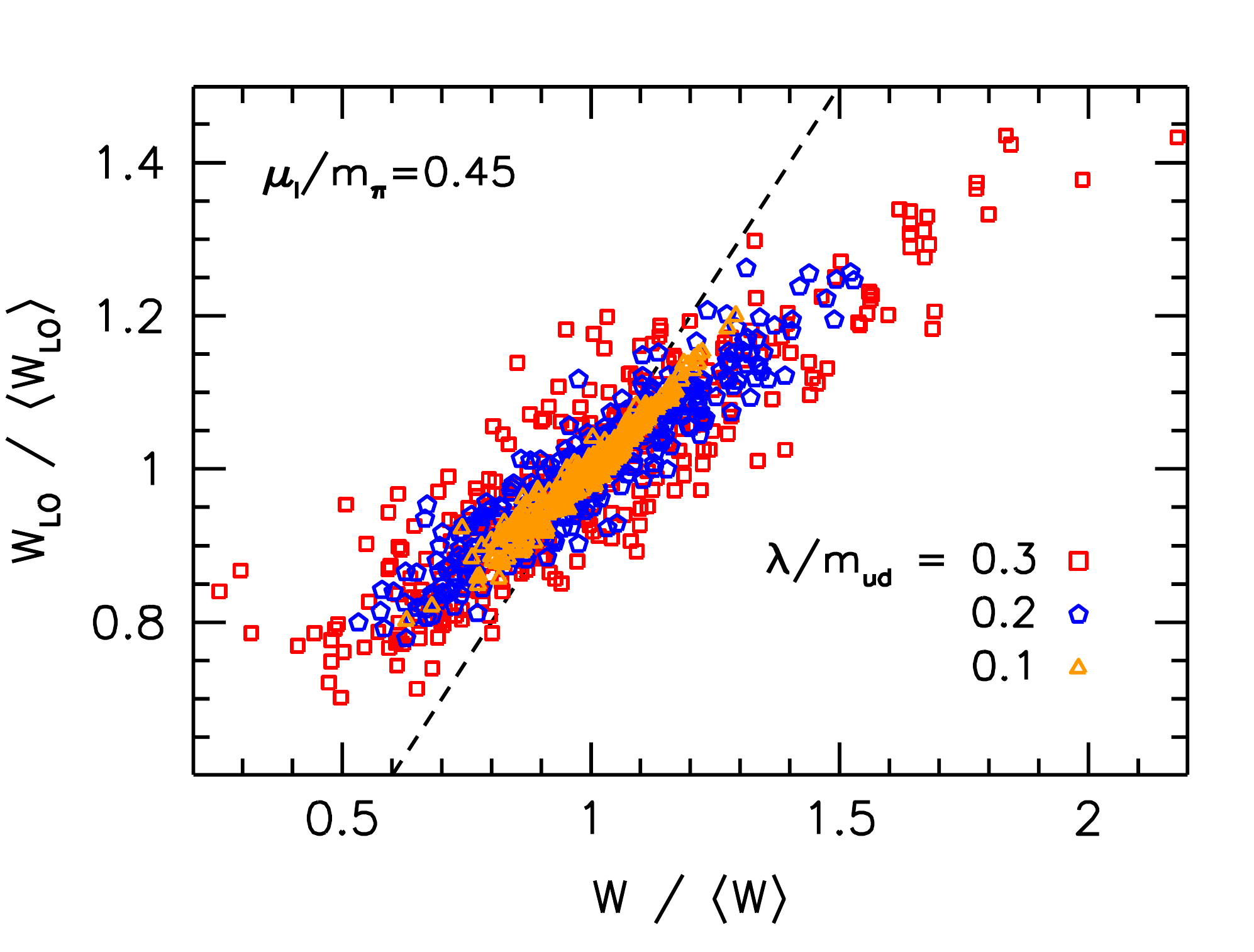}\quad
 \includegraphics[width=8cm]{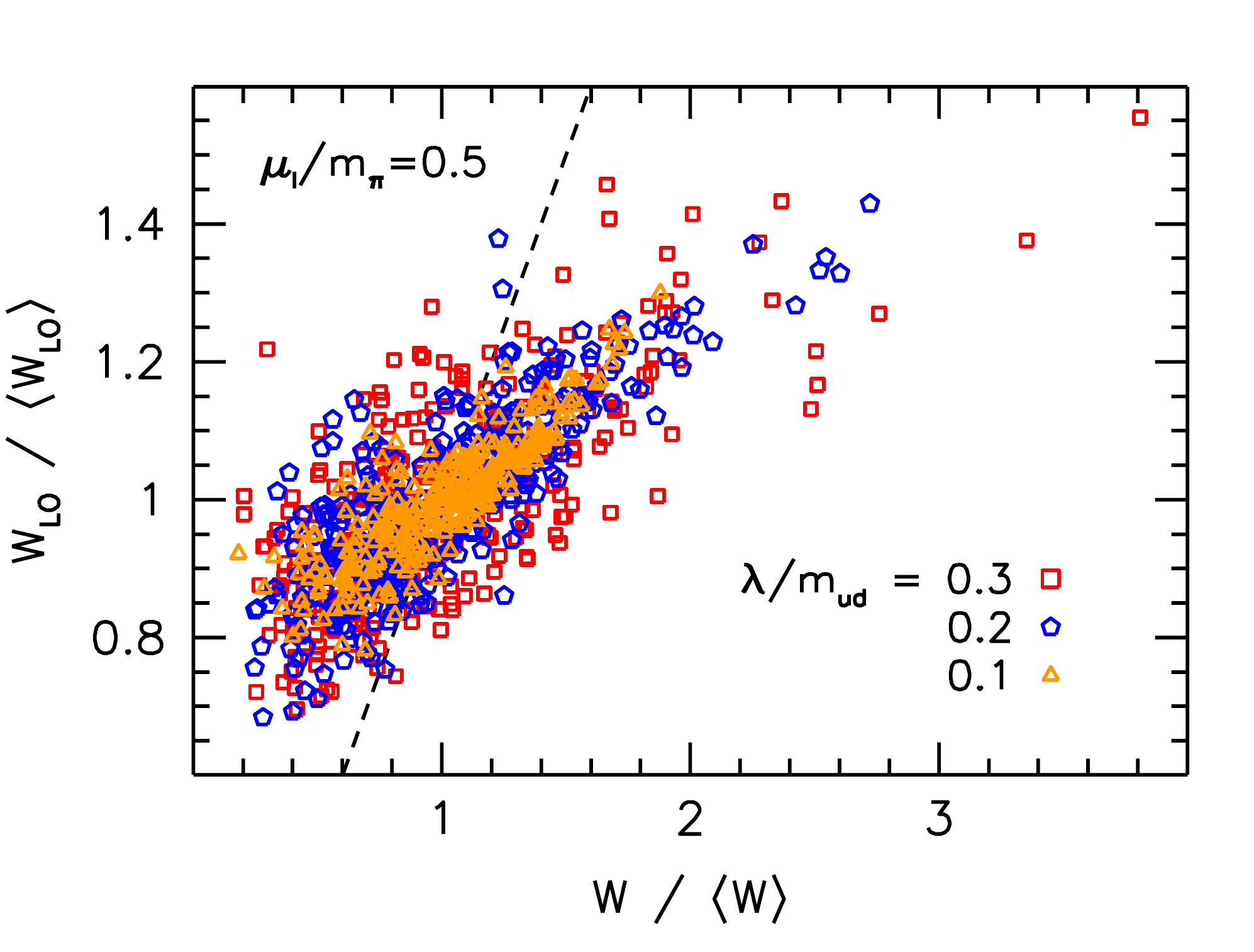}
 \caption{\label{fig:scatterW}The leading-order reweighting factor $W_{\rm LO}$ 
 against the full weight $W$ on a scatter plot for two isospin chemical potentials 
 at low temperature. The dashed line represents $W_{\rm LO}=W$. 
For this plot the $8^4$ ensemble of Ref.~\cite{Endrodi:2014lja} was 
 used. }
\end{figure}

Thus we conclude that the reweighting of an observable, to leading order in $\lambda$, 
involves the exponential of the pion condensate (measured at $\lambda>0$) 
times the four-volume. Since the pion 
condensate is anyway measured on the individual configurations to compute $\expv{\pion}$,
this improvement comes with no extra costs. 
We test this improvement on small lattices, where a calculation of the complete spectrum 
of singular values $\xi$ is feasible, enabling 
a direct comparison between $W$ and $W_{\rm LO}$. 
Specifically, we employ the low-temperature 
$8^4$ ensembles generated in Ref.~\cite{Endrodi:2014lja} and 
plot the two reweighting factors against each other in Fig.~\ref{fig:scatterW} for 
two values of $\mu_I$ around the critical isospin chemical potential. 
The scatter plot clearly shows the strong correlation between $W$ and $W_{\rm LO}$ 
and that the two factors become identical in the limit
$\lambda\to0$.\footnote{Notice furthermore that $W_{\rm LO}$ tends to overestimate $W$
if the reweighting factors are below average (and vice versa). 
This can be understood using the singular value 
representation of Eqs.~(\ref{eq:rewe}) and~(\ref{eq:LOrewe}),
\be
\log W = \frac{1}{4}\sum_n \log \frac{\xi_n^2}{\xi_n^2+\lambda^2}, \quad\quad\quad
\log W_{\rm LO} = \frac{1}{4} \sum_n \frac{-\lambda^2}{\xi_n^2+\lambda^2}\,,
\ee
which can be used to show that the difference
$W_{\rm LO}-W$ is a positive and monotonously decreasing function of $\xi_n$ at fixed $\lambda$.
Thus, a fluctuation that reduces the singular values (i.e.\ that pushes $W$ below average), 
inevitably increases the deviation $W_{\rm LO}-W$, thereby explaining the tendency visible 
in Fig.~\ref{fig:scatterW}.}

\section{Results: improvement program}
\label{sec:results}

In the following we always perform the leading order reweighting with $W_{\rm LO}=\exp[-\lambda V \pion/(2T)]$,
where $\pion$ is calculated from~(\ref{eq:traces1a}) employing
noisy estimators to evaluate the traces. The measurements are carried out on the reweighted ensembles.

\subsection{Improved pion condensate}
\label{sec:rho0-extract}

\begin{figure}[ht!]
 \centering
 \includegraphics[width=8cm]{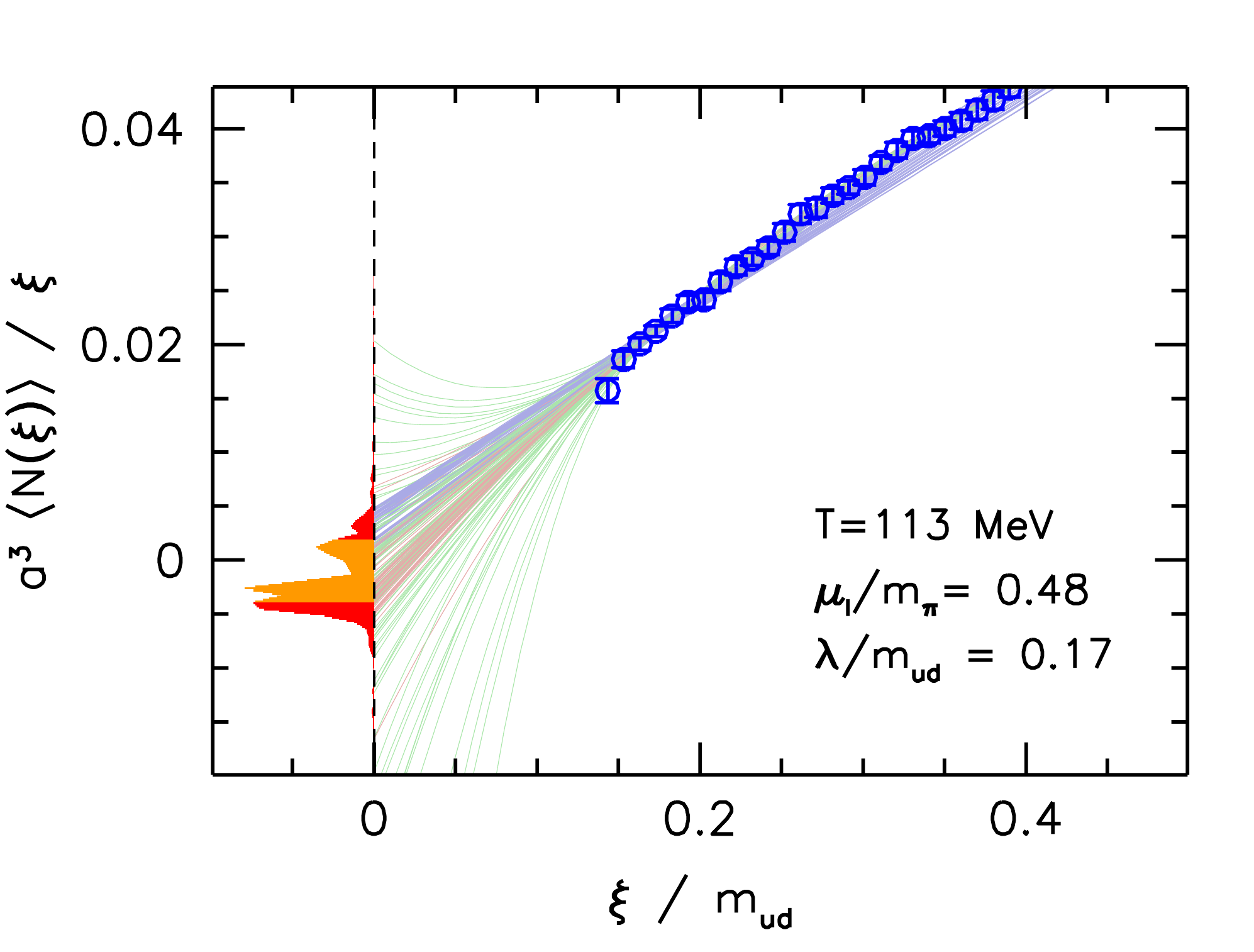}\quad
 \includegraphics[width=8cm]{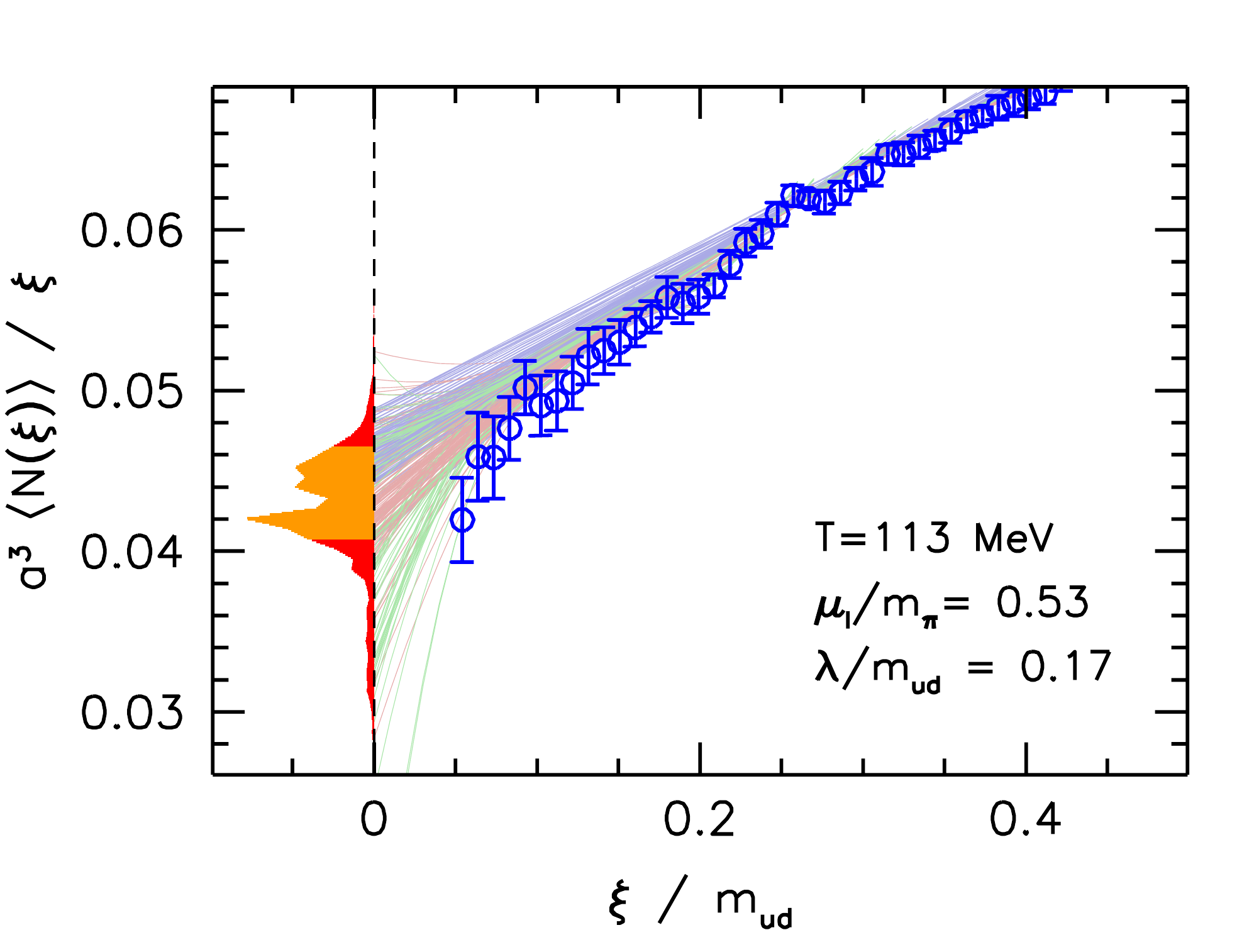}
 \caption{\label{fig:rho0fit}
 Integrated spectral density (blue points) 
 of the singular values of $\Dp+\ml$, as measured on our $24^3\times6$ 
 ensemble at $T=113 \textmd{ MeV}$ for isospin chemical potentials just
 around (top panel) 
 and slightly beyond (bottom panel) the transition to the BEC phase. The
 colored curves indicate polynomial 
 fits of the data (blue: linear, red: quadratic, green: cubic) 
 and the red vertical histograms mark the distribution of $\rho(0)$ as obtained from the 
 fits. The orange part of the histogram is our estimate of the statistical plus 
 systematical error of the fits. 
 }
\end{figure}

We describe our strategy to determine $\rho(0)$ in more detail. The singular values of the 
massive Dirac operator (i.e., the eigenvalues of $|\Dp+\ml|^2$, cf.\ Eq.~(\ref{eq:sval})) are calculated using the Krylov-Schur
algorithm. We find it sufficient to work with the lowest 50-150 singular values for each configuration. 
Using this set of singular values we build a histogram for the integrated spectral density
\be
N(\xi) = \int_0^\xi \dd \xi' \rho(\xi')\,,
\ee
where we discard bins that lie below the average lowest singular value. 
The statistical error of $N(\xi)$ in each bin is estimated via the jackknife
procedure. In Fig.~\ref{fig:rho0fit} we plot $N(\xi)/\xi$ 
for two ensembles just around and slightly beyond the transition to the pion 
condensed phase. 

To obtain $\rho(0)$, we need to extrapolate $N(\xi)/\xi$ down to zero. This is performed via 
polynomial fits of the histogram. To take into account the correlation between the individual 
bins, we minimize the correlated $\chi^2_{\rm corr}$, 
which involves the inverse of the correlation matrix $C$ 
of the data. To avoid numerical problems during this inversion, we smear the lowest eigenvalues of $C$
following the strategy of Ref.~\cite{Michael:1994sz}. We perform the fits with polynomials of 
various degrees and over various fit ranges. The resulting values for $\rho(0)$ 
are weighted by $\exp(-\chi^2_{\rm corr}/N_{\rm dof})$ with $N_{\rm dof}$
the number of degrees of freedom in the fit. The weighted results are used to build yet another histogram
(red area in Fig.~\ref{fig:rho0fit}), similar 
to the analysis conducted in Ref.~\cite{Borsanyi:2014jba}. The 
central value for $\rho(0)$ is obtained by the median, while the 
statistical plus systematical error of the fit is estimated by the range that contains
the middle $68\%$ of the 
histogram (orange area in Fig.~\ref{fig:rho0fit}). 
The analysis gives drastically different results for the two chemical 
potentials considered in the figure.
While the intersect of the fits is large for $\mu_I/m_\pi=0.53$, it is 
consistent with zero for $\mu_I/m_\pi=0.48$. 
(In the latter case $\rho(0)$ and its error is taken as the positive part of the 
orange histogram.)
This demonstrates -- via the relation~(\ref{eq:BC}) -- 
the emergence of a nonzero 
pion condensate  in the BEC phase.
The so obtained results for $\expv{\pion}$ are included in 
the top panel of Fig.~\ref{fig:pbGp_impro} above. 

\subsection{Improved quark condensate}

In principle, for the improvement scheme of the quark condensate, outlined in Sec.~\ref{sec:obsimpr},
we need all singular
values $\xi_n$. This is in contrast to the improvement of the pion condensate, where 
only the spectral density of singular values around zero is of importance. Nevertheless, computing all
singular values for each configuration is unfeasible in practice and, in fact, not necessary. The important
simplification follows from the observation that the difference $\delta_{\bar\psi\psi}$ 
from Eq.~(\ref{eq:ldiffs}) is dominated by the $\lambda$ dependence of the terms involving the lowest
singular values. It might thus be sufficient to perform the explicit $\lambda\to0$ extrapolation in the
operator for only the first $\Nxi$ low modes and to leave the contribution from the remaining singular
values untouched. To this end we introduce the truncated operator
\be
\bar\psi\psi^{\Nxi}(\lambda) = \frac{T}{2V} \sum_{n=1}^{\Nxi} \frac{\textmd{Re}\;
\varphi_n^\dagger [\Dp+\ml] \varphi_n}{\xi^2_n+\lambda^2}\,,
\ee
and the associated truncated difference
\be
\label{eq:ldiffs-trunc}
\begin{split}
\deltaNxi_{\bar\psi\psi} &\equiv \bar\psi\psi^{\Nxi}(\lambda)-\bar\psi\psi^{\Nxi}(\lambda=0) \\
&= 
\frac{T}{2V} \sum_{n=1}^{\Nxi} \textmd{Re}\; \varphi_n^\dagger [\Dp+\ml] \varphi_n \cdot \left(
\frac{1}{\xi_n^2+\lambda^2}-\frac{1}{\xi_n^2}\right)\,,
\end{split}
\ee
to rewrite the expectation value as
\be
\label{eq:impr-expv}
\expv{\bar\psi\psi}=\expv{\bar\psi\psi-\deltaNxi_{\bar\psi\psi}} + \expv{\deltaNxi_{\bar\psi\psi}} \,.
\ee
The second term vanishes when we perform the $\lambda$-extrapolation, so that we obtain
\be
\label{eq:liml0}
\lim_{\lambda\to0} \expv{\bar\psi\psi}=\lim_{\lambda\to0} \expv{\bar\psi\psi-\deltaNxi_{\bar\psi\psi}} \,.
\ee
This type of improvement will be efficient if $\Nxi$ is large enough to ensure that
$\deltaNxi_{\bar\psi\psi}\approx\delta_{\bar\psi\psi}$ so that the subtraction effectively removes
most of the $\lambda$ dependence of the observable.

\begin{figure}[t]
 \centering
 \includegraphics[width=8cm]{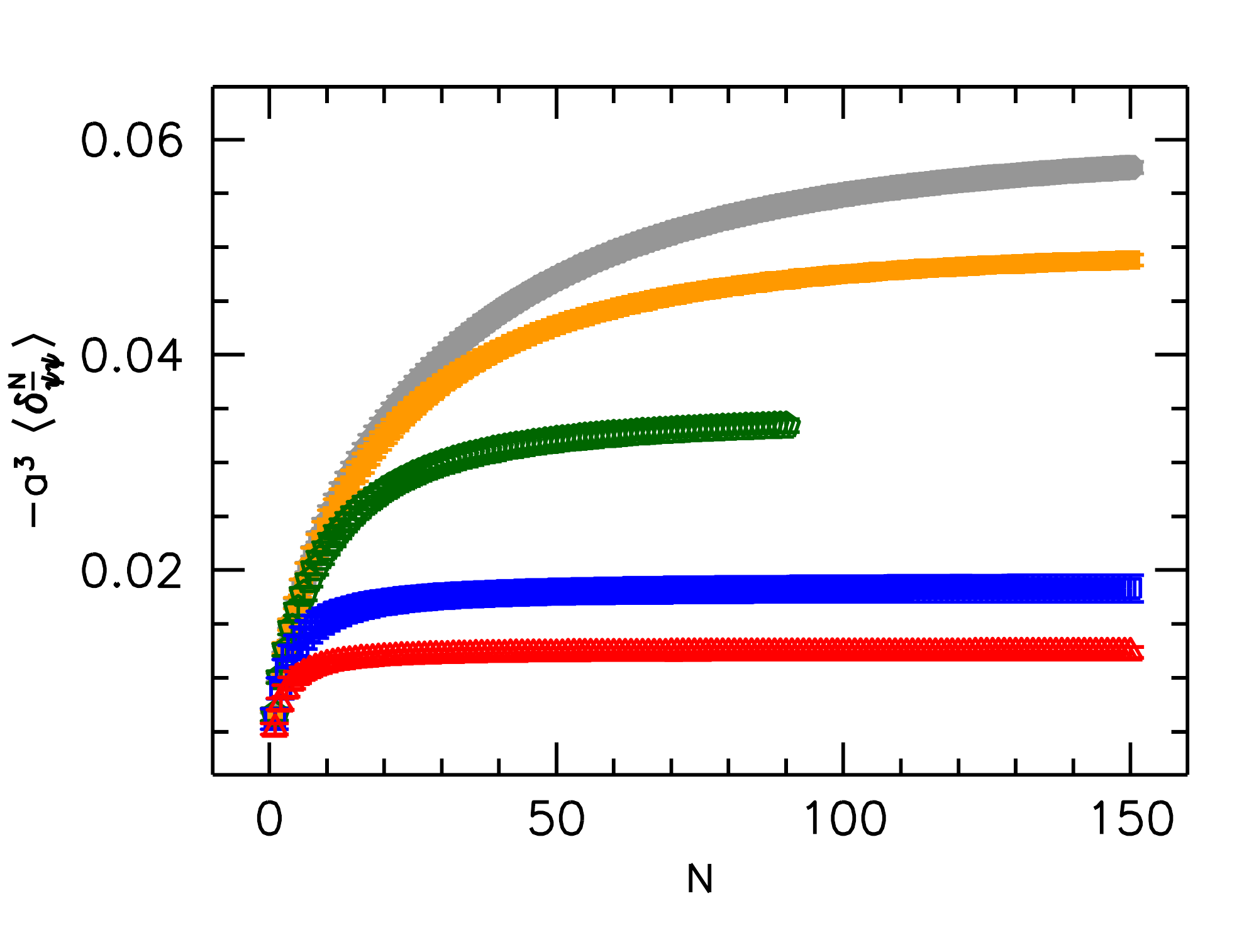}\quad
 \includegraphics[width=8cm]{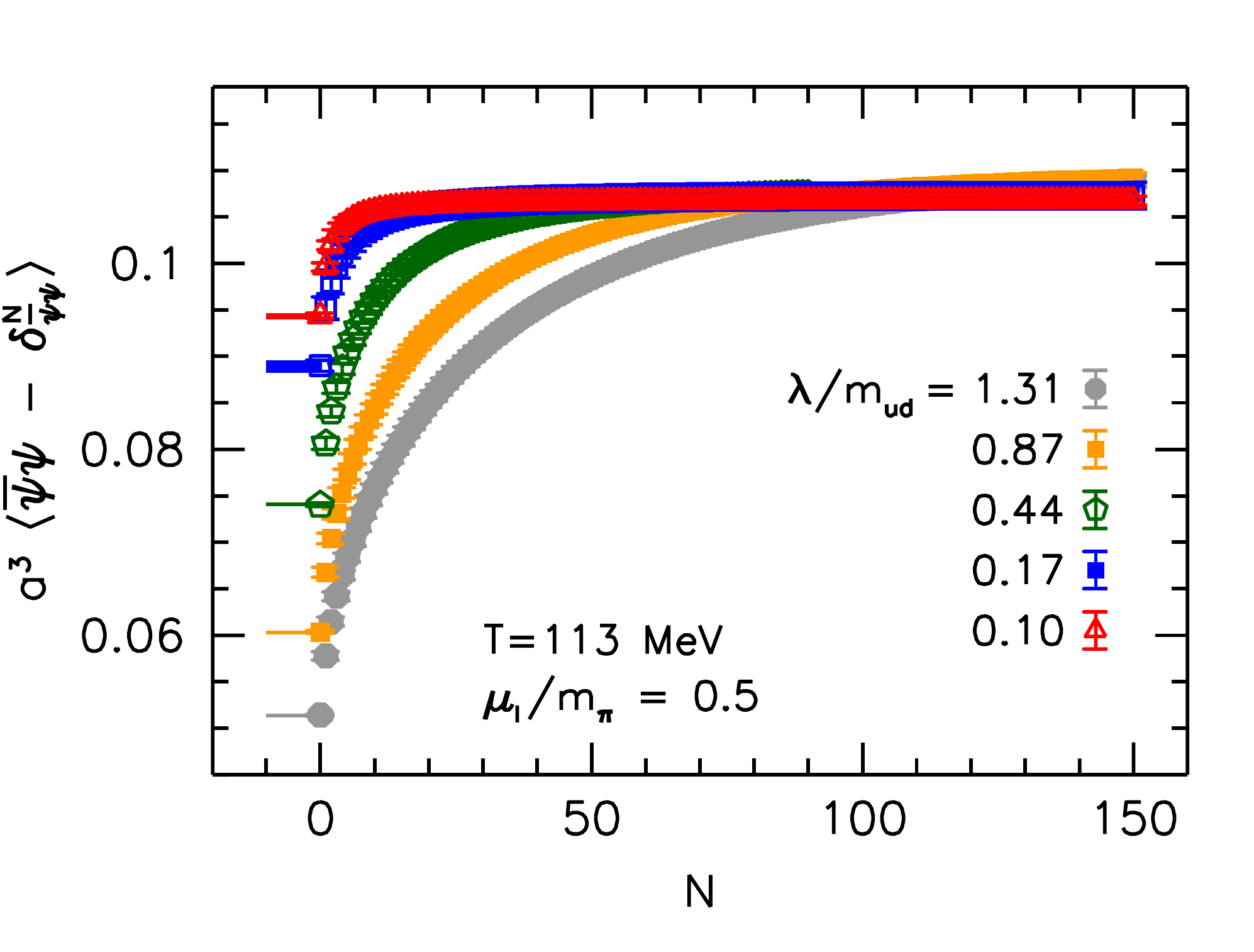}
 \caption{\label{fig:pbp_impro_sv}
 Top panel: the difference $\delta_{\bar\psi\psi}^N$, calculated using the $N$ lowest singular values, 
 on ensembles with various values of $\lambda$. Bottom panel: the improved condensate~(\ref{eq:impr-expv}) 
 for the same ensembles. The horizontal lines indicate the results for $\expv{\bar\psi\psi}$ obtained 
 using noisy estimators of the trace~(\ref{eq:traces1a}). The color coding is the same in both panels. 
 }
\end{figure}

The optimal value for $\Nxi$ to achieve a good balance between the improvement for the
$\lambda$ dependence of the observable and the computational cost will, in general, depend
on the operator, the lattice size and the lattice spacing, as well as on the magnitude of
the $\lambda$-values in use.  The results for $\deltaNxi_{\bar\psi\psi}$
versus $\Nxi$ for different values of $\lambda$ are shown in the top panel of Fig.~\ref{fig:pbp_impro_sv}. The
figure indicates that moderate values of $N$ suffice to ensure that most
of the $\lambda$ dependence in the valence sector is absorbed by the improvement. 
In the bottom panel of Fig.~\ref{fig:pbp_impro_sv} we plot the combination
$\expv{\bar\psi\psi-\deltaNxi_{\bar\psi\psi}}$, revealing no significant $\lambda$ dependence for $N\gtrsim100$. 
For comparison, the $\lambda$ dependence of the unimproved observable (marked by the
short horizontal lines in the figure) is much more pronounced.

We find that $\Nxi\sim100$ does suffice to warrant well controlled
$\lambda$-extrapolations for all ensembles and observables considered so far. 
Thus, the same set of singular values can be used for the improvement of the pion condensate 
as well as for the quark condensate.

\subsection{Final $\lambda$-extrapolations}

\begin{figure}[ht!]
 \centering
 \includegraphics[width=8cm]{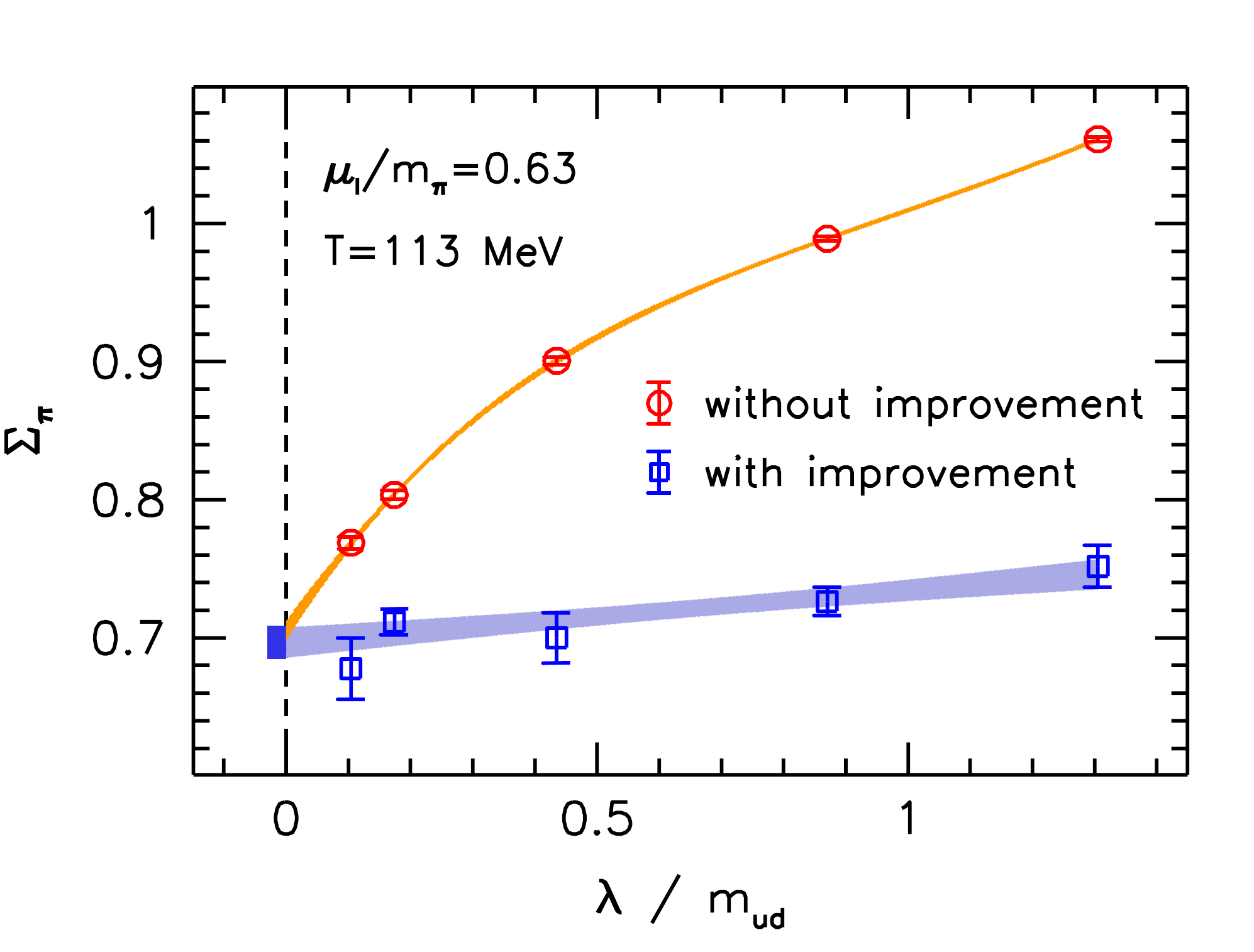}
 \includegraphics[width=8cm]{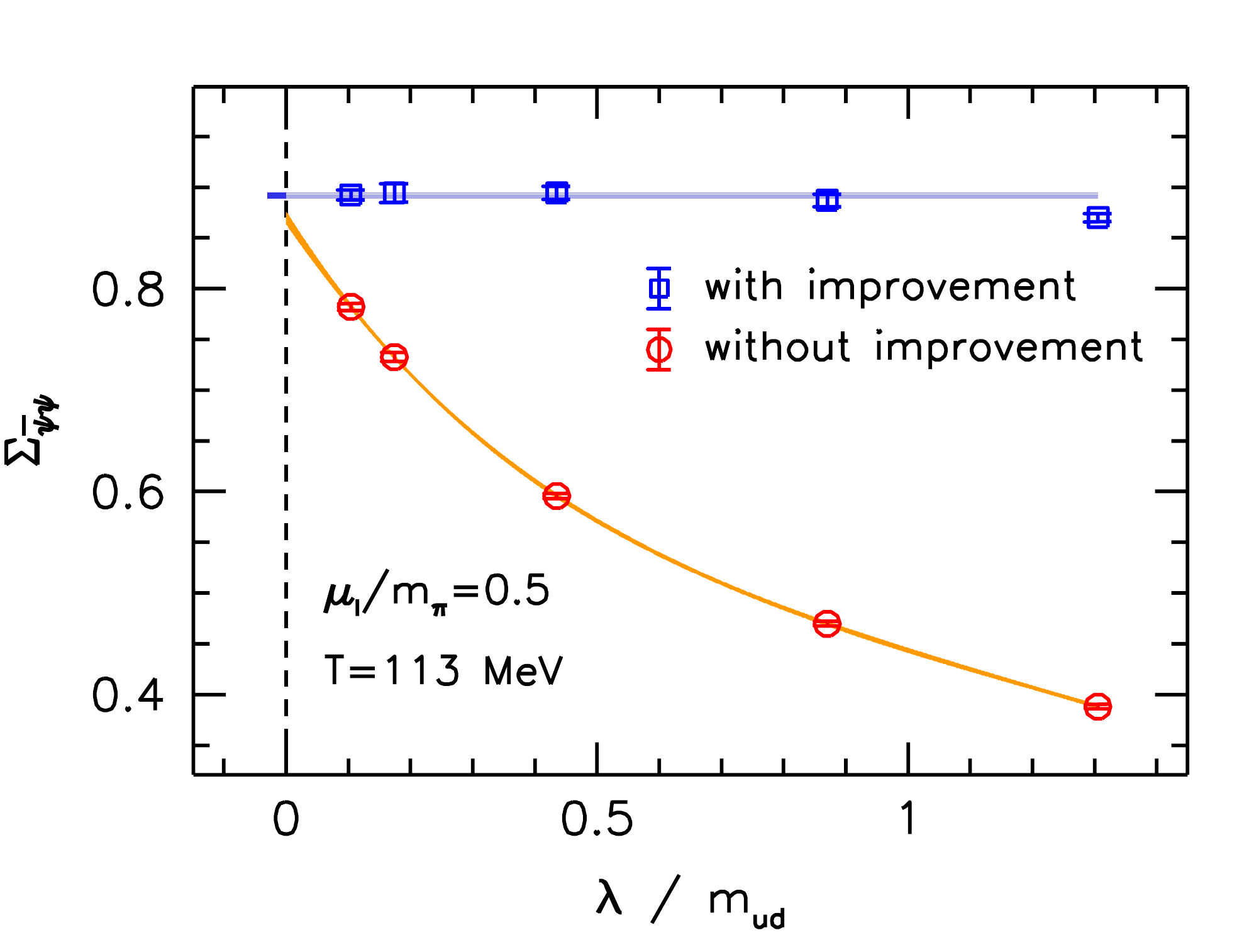}
 \caption{\label{fig:pbpimpro}
 Final $\lambda$-extrapolations for the pion (top panel) and quark condensates (bottom panel) 
 on our $24^3\times 6$ ensembles 
 for various values of the isospin chemical potential. 
 The improvement is crucial in all cases and enables a reliable extrapolation to $\lambda=0$.
 }
\end{figure}

Having performed the improvements described above, we can carry out the final extrapolation of the 
results to $\lambda=0$. 
Since the improvements do not achieve a perfect elimination of the pionic source [the reweighting is 
done only at leading order, the singular value sums are truncated at a finite $N$, and the finiteness 
of the volume distorts the relation~(\ref{eq:BC})], a mild dependence of the results on $\lambda$ 
is still expected. In Fig.~\ref{fig:pbpimpro} we show a few representative examples
for the extrapolations of $\Sigma_\pi$ and of $\Sigma_{\bar\psi\psi}$, where we also 
include the unimproved observables [i.e. the quantities obtained via the full traces, Eq.~(\ref{eq:traces1a})].
Contrary to the pronounced (and thus uncontrolled) 
dependence of the latter on $\lambda$, we find that the improved observables 
can be reliably fitted by either a constant or a linear function (on general grounds, $\Sigma_{\pi}$ is 
an odd function of $\lambda$, whereas $\Sigma_{\bar\psi\psi}$ is even in the pionic source).
We also attempted to perform a fit of the unimproved data by means of a spline with Monte-Carlo-generated 
nodepoints (for the details of this fit procedure, see Ref.~\cite{Brandt:2016zdy}). For comparison, 
the so obtained extrapolations are also included in Fig.~\ref{fig:pbpimpro}. Unlike the improved 
extrapolations, these fits are always dominated by the points at low $\lambda$, making them unreliable
in several cases.

\section{Results}

\subsection{Phase diagram}

\begin{figure}[t]
 \centering
 \includegraphics[width=9.1cm]{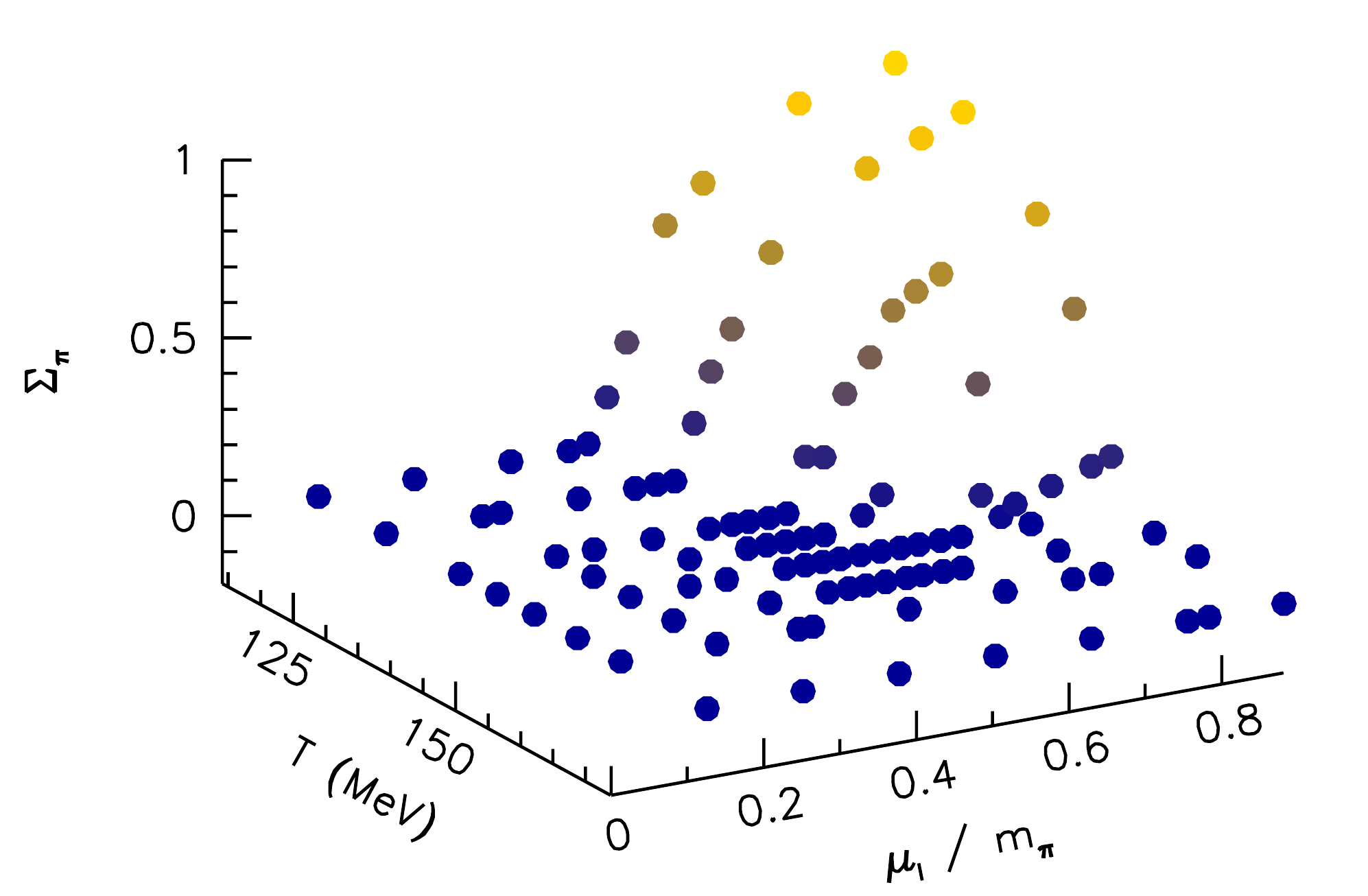}
 \includegraphics[width=9.1cm]{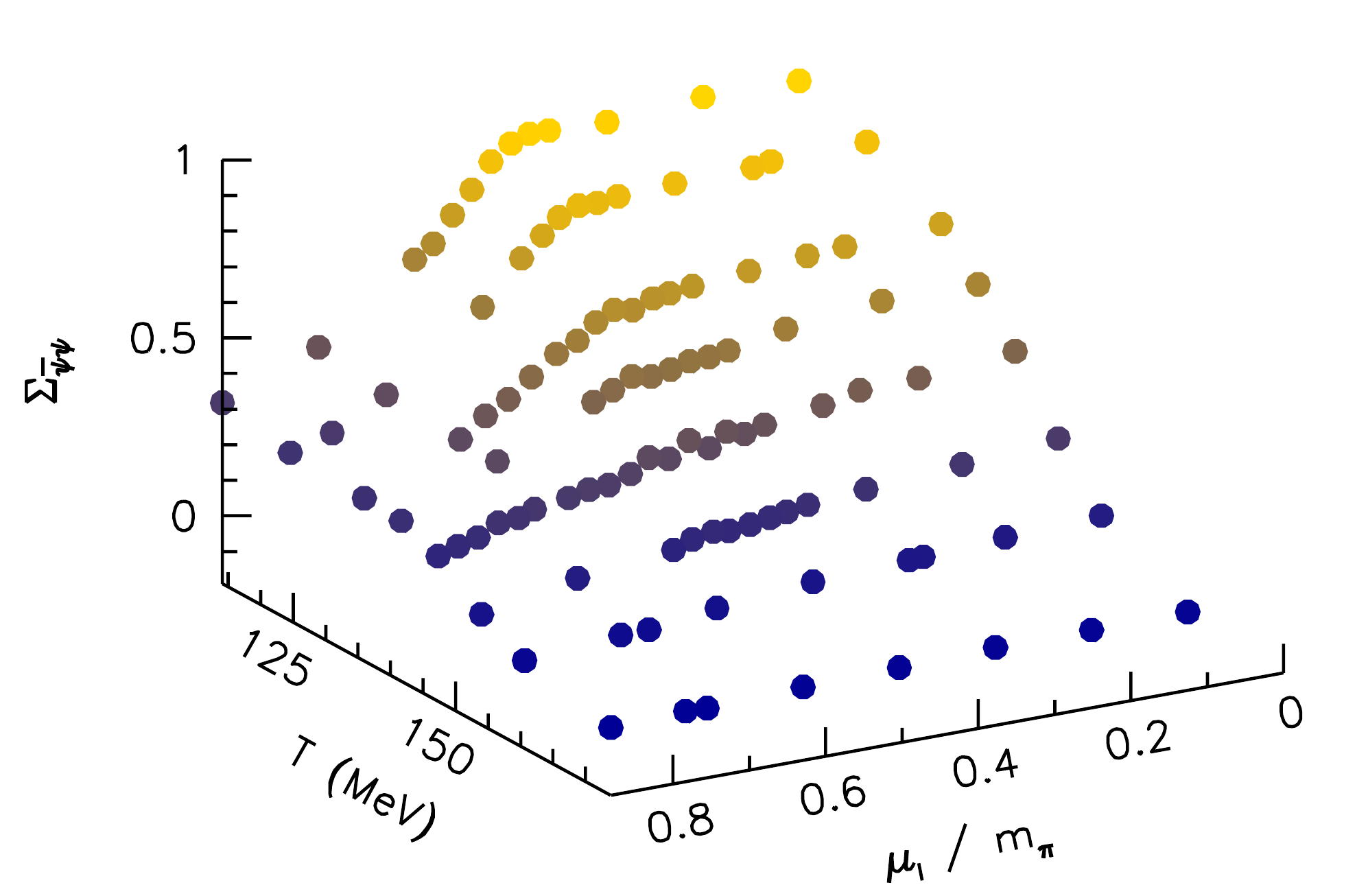}
 \caption{\label{fig:pbp_surface}
 The pion condensate (top panel) and the quark condensate (bottom panel) as 
 functions of the temperature and the isospin chemical potential
 as measured on our $N_t=6$ lattices. The color coding reflects the magnitude of 
 the observables: blue ($\Sigma=0$) towards yellow ($\Sigma=1)$. Notice that the 
 orientation of the $\mu_I$-axis is different in the two panels. 
 }
\end{figure}

In the following we work with the $\lambda\to0$ extrapolated observables to determine the phase diagram of
the system in the $\mu_I-T$ plane. We work with four lattice ensembles with $N_t=6$, $8$, $10$ 
and $12$.
Fig.~\ref{fig:pbp_surface} shows our $N_t=6$ results for $\Sigma_\pi$
and for $\Sigma_{\bar\psi\psi}$ from Eq.~(\ref{eq:renobs}). 
The onset of pion condensation is characterized by the abrupt change at the boundary between the vacuum and 
the BEC phase, mapping a critical line $\mu_{I,c}(T)$. 
The quark condensate exhibits a similarly pronounced change at the pion condensation phase boundary. 
In addition, it also features a smooth dependence on the temperature around the chiral crossover
transition $\Tc(\mu_I)$. In the following, we refer to the point,
where these two transition lines meet as the
{\it pseudo-triple point} at $\mu_I=\mut$ and $T=\Tt$. 
The notation refers to the fact that the chiral transition is pseudocritical and 
does not correspond to a true phase transition (in which case this point would be a true triple point).

The boundary of the pion condensation phase, $\mu_{I,c}(T)$, is determined by the points where $\Sigma_\pi$ acquires a nonzero
expectation value. We estimate its location by interpolating\footnote{A technical issue regarding this fit is the non-analyticity of $\Sigma_\pi$ at the 
critical chemical potential. Although it is regulated by the finite volume, the sharp behavior
around $\mu_{I,c}$ cannot be captured by the smooth splines. To avoid this issue, here 
we fit the results from Sec.~\ref{sec:rho0-extract}, allowing for negative
intersects for $\rho(0)$ (i.e.\ the full orange histograms of Fig.~\ref{fig:rho0fit}).
This enables a more stable determination of the $\Sigma_\pi=0$ contour. } the results for $\Sigma_\pi$ using
a two-dimensional cubic spline fit with Monte-Carlo-generated nodepoints~\cite{Brandt:2016zdy}.
From $\chi$PT it is known that $\mu_{I,c}(0)=m_\pi/2$~\cite{Son:2000xc}. We observe that $\mu_{I,c}(T)$ is independent of $T$ up to a temperature
of about $130$ to $140$~MeV -- see also the results shown in~\cite{Brandt:2016zdy,Brandt:2017zck}.
For a temperature of about $155$ to $160$~MeV the phase boundary
starts to flatten out, and above $T\approx161$~MeV the pion condensate is
found to vanish within errors for all ensembles.
Here we follow the phase boundary up to $\mu_I=120$~MeV, but this flat behavior 
is found to persist at least up to $\mu_I\approx190$~MeV on our $N_t=6$ ensemble, see below.
These findings contradict
the expectations from $\chi$PT, which predicts a monotonous rise 
in $\mu_{I,c}(T)$~\cite{Son:2000xc}. 

\begin{figure}[t]
 \centering
 \includegraphics[width=8cm]{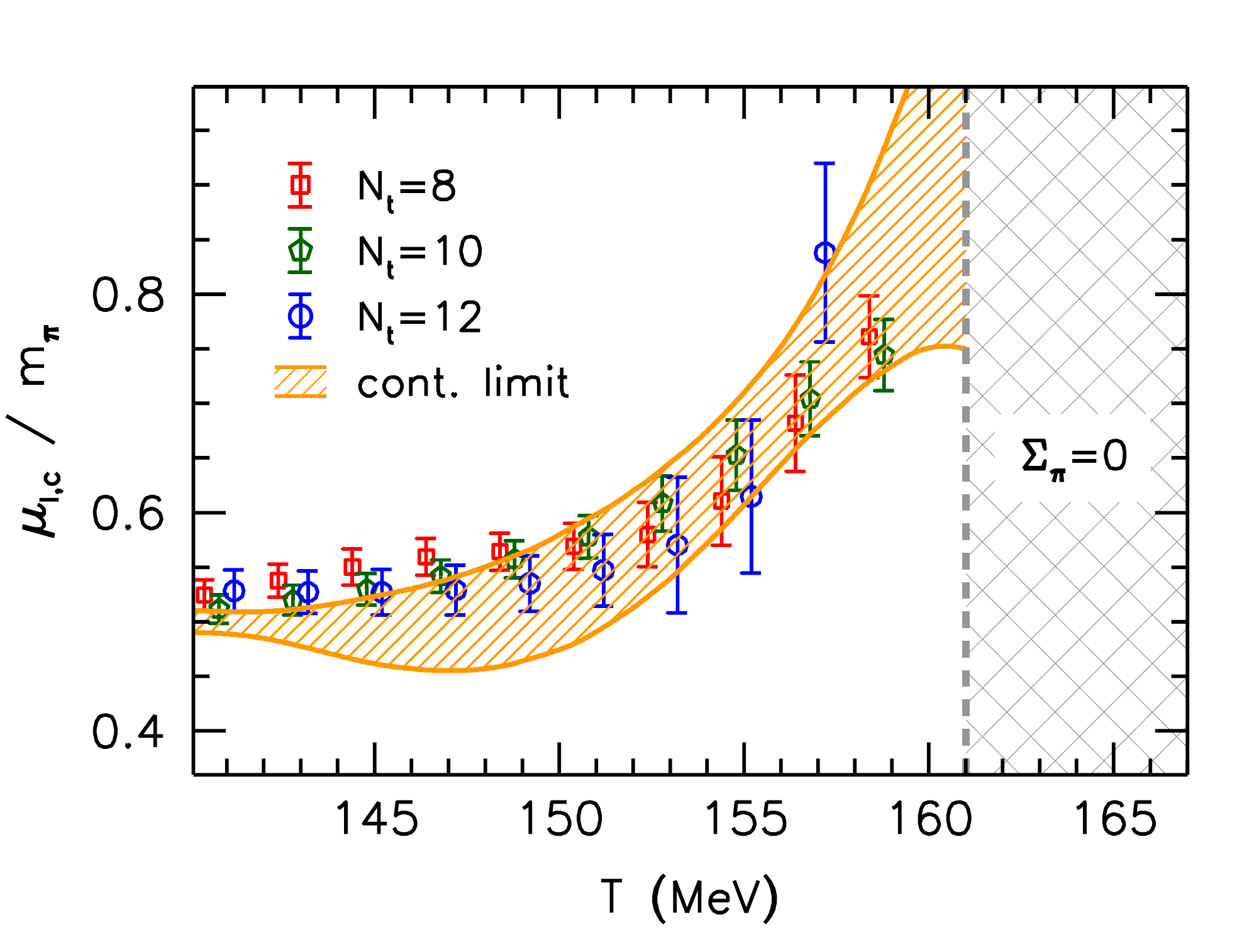}
 \includegraphics[width=8cm]{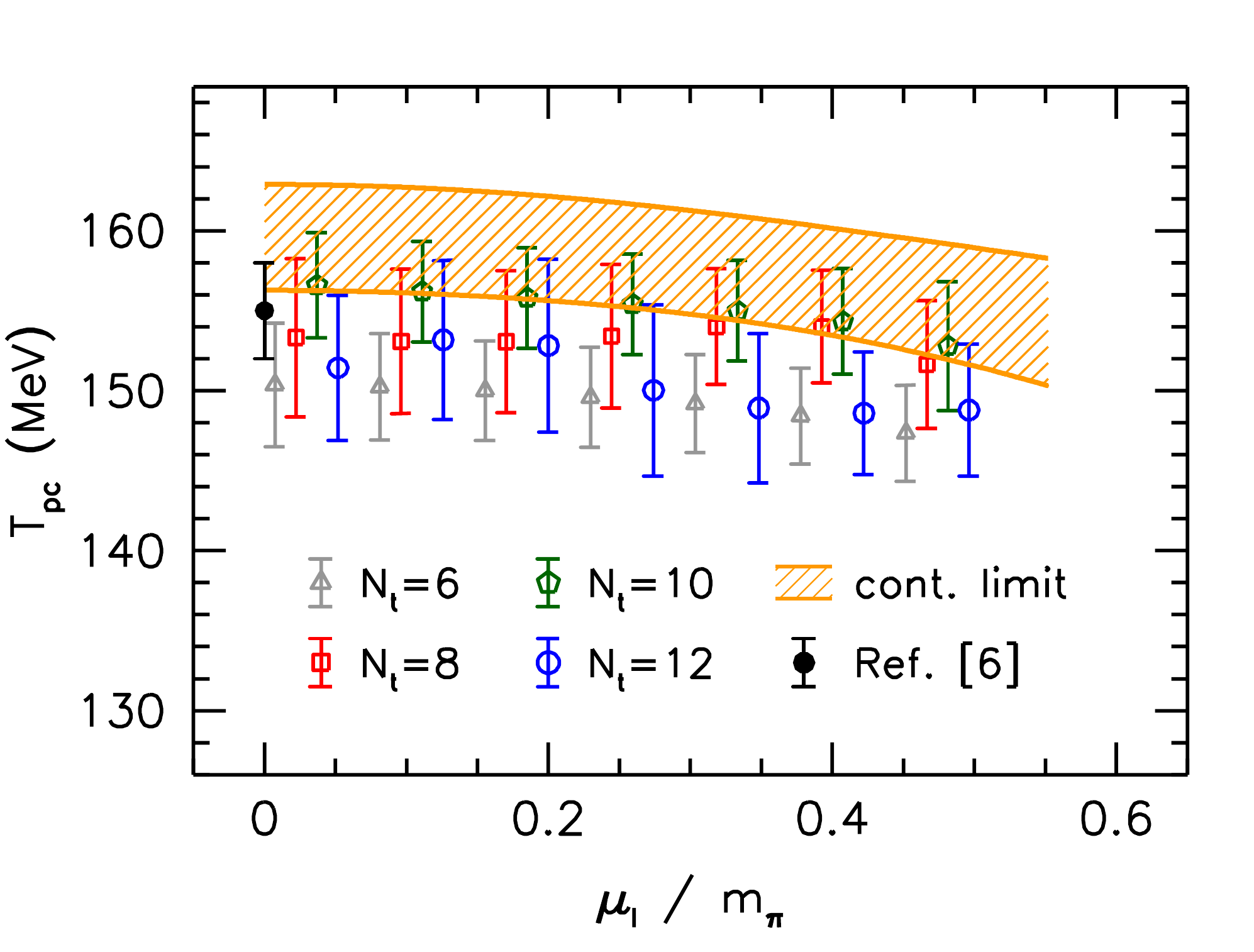}
 \caption{\label{fig:conti-extra}
 Continuum extrapolations for the pion condensation phase boundary (top panel) and the chiral crossover transition
 temperature (bottom panel). The yellow curves correspond to the continuum extrapolations, and we also show
 the data for the individual lattices which have been included in the fits. 
 In the upper panel, the shaded gray area corresponds to the region where $\Sigma_\pi$ is 
 observed to be consistent with zero within errors -- serving as an upper bound for the 
 critical temperature along the pion condensation boundary.
 }
\end{figure}

To perform the continuum extrapolation, we
parameterize $\mu_{I,c}(T,a)$ by a polynomial in $(T-T_0)$ with lattice spacing dependent coefficients. We set
$T_0=140$~MeV, below which $\mu_{I,c}(T,a)$ is found to be independent of $T$. To capture 
both the flatness of the critical chemical potential for low temperatures and 
its abrupt rise around $T\approx 155 \textmd{ MeV}$, 
we found it necessary
to include terms proportional to $(T-T_0)^2$,
$(T-T_0)^3$ and $(T-T_0)^4$ in the fit and 
to fix $\mu_{I,c}(T_0,0)=m_\pi/2$.
We find that for $140\textmd{ MeV}\lesssim T \lesssim 155\textmd{ MeV}$, our $N_t=6$ results
are outside of the
scaling region for lattice artefacts of $\O(a^2)$ so that we excluded these from the continuum extrapolation. The final
continuum curve, extrapolated including terms proportional to $a^2$, is shown in Fig.~\ref{fig:conti-extra} (top panel),
together with the data for the phase boundary which has been included in the fit. 
$T\approx161$~MeV serves as an upper bound for the critical temperature, as 
$\Sigma_\pi$ is found to vanish within the uncertainties for all lattice spacings and for all
values of $\mu_I$ considered in this region, as mentioned above. 

We proceed with the chiral crossover transition temperature $\Tc(\mu_I)$, which 
we define as the location of the inflection point of
$\Sigma_{\bar\psi\psi}$ with respect to $T$.\footnote{Note
that in Refs.~\cite{Brandt:2016zdy,Brandt:2017zck} we employed a different procedure, 
defining $\Tc$ as the
point where $\Sigma_{\bar\psi\psi}(\mu_I,\Tc(\mu_I))=\Sigma_{\bar\psi\psi}(0,\Tc(0))$ holds.
This definition is only valid for $\mu_I<\mu_{I,c}(0)$, since beyond that the condensate
is affected by $\mu_I$ already at $T=0$. 
On the contrary, the inflection point can be
used to define $\Tc(\mu_I)$ at any value of $\mu_I$.} 
The data shows an initial reduction in $\Tc(\mu_I)$ up to about $\mu_I=70\textmd{ MeV}$, 
followed by a slight increase. Beyond the pseudo-triple point, where the crossover line and 
the pion condensation boundary meet, we observe the two transition lines to coincide within 
errors. To be more quantitative, we once again determine $\Tc(\mu_I)$ using a two-dimensional
spline fit where the nodepoints have been generated via Monte-Carlo. In the spline fit\footnote{
Here we only take into account the fit for $T>135\textmd{ MeV}$, where the spline 
captures the behavior of $\Sigma_{\bar\psi\psi}(\mu_I,T)$ sufficiently accurately. 
Below this temperature the condensate changes abruptly at the BEC phase transition, which 
is not captured by our smooth interpolation.
}
we include the constraint that
$\Sigma_{\bar\psi\psi}$ is an even function of $\mu_I$.
To perform the continuum
extrapolation, we parameterize $\Tc(\mu_I,a)$ by a polynomial which is even in $\mu_I$,
including data up to $\mu_{I,c}(0)=m_\pi/2$.
The behavior of $\Tc(\mu_I)$ beyond this value is difficult to capture with polynomials. 
For the range of chemical potentials included in the fit,
the data is already well described by a quadratic polynomial in $\mu_I$ 
and $\O(a^2)$ lattice artefacts,
see Fig.~\ref{fig:conti-extra} (bottom panel). Adding a $\mu_I^4$-term or higher-order lattice 
discretization errors was found to only increase the uncertainties and not lead to 
significant differences.

\begin{figure}[t]
 \centering
 \includegraphics[width=8cm]{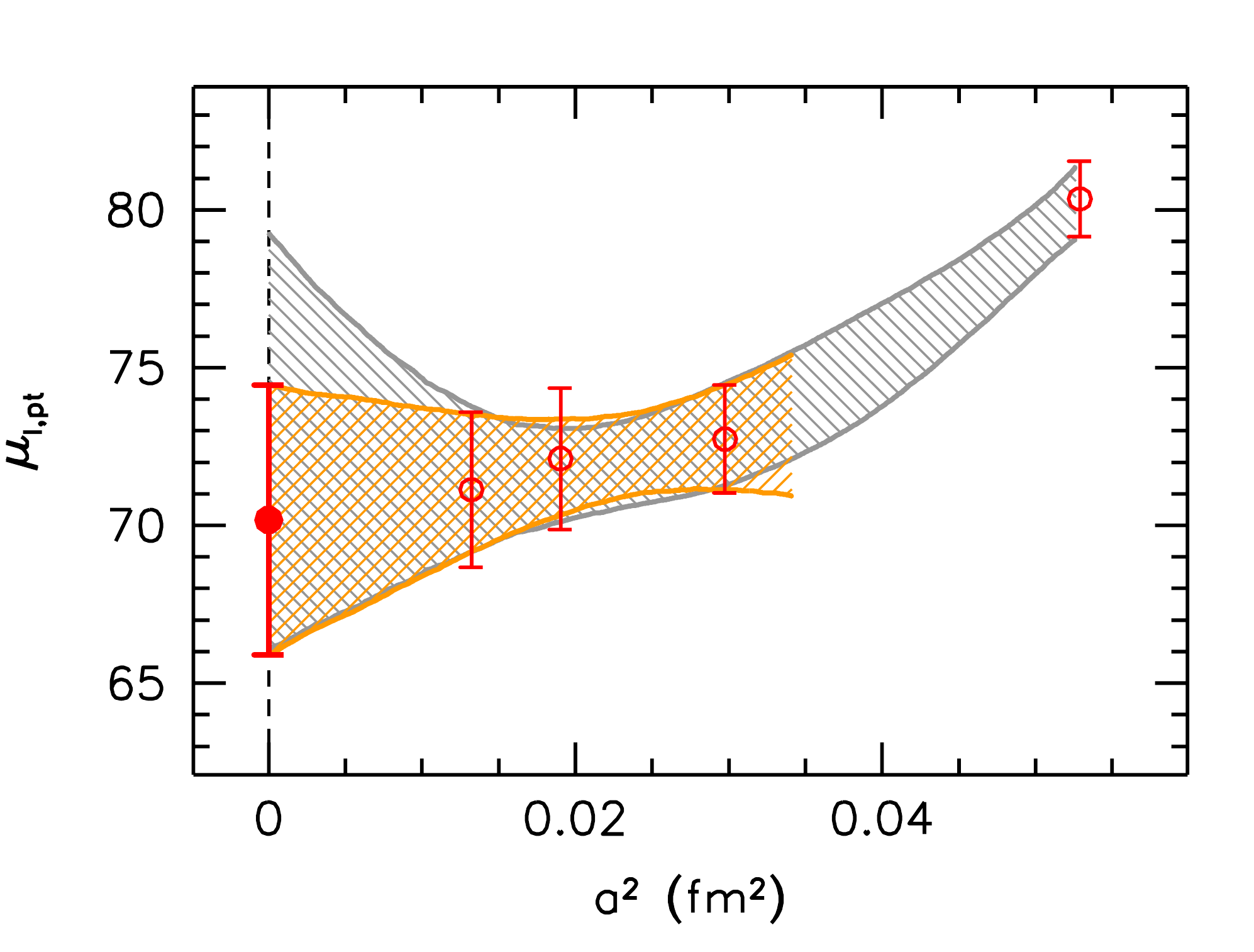}
 \caption{\label{fig:mp-conti-extra}
 Continuum extrapolations for the isospin chemical potential $\mut$ 
 corresponding to the pseudo-triple point, where the chiral crossover line meets the pion 
 condensation boundary. The orange curve corresponds to the continuum extrapolation
 for $N_t=8,\,10$ and 12 including lattice artefacts of $O(a^2)$ and the gray curve
 is the continuum extrapolation for all points including an additional $O(a^4)$ term. 
 }
\end{figure}

The precise location of the pseudo-triple point is also of interest. 
We define $\mut(a)$ to be the value for $\mu_I$ where the two transition lines become consistent within errors -- i.e.\ where
the uncertainties overlap. This provides a conservative lower bound for the pseudo-triple point.
The results
for $\mut(a)$ are shown in Fig.~\ref{fig:mp-conti-extra}. As for the pion condensation phase boundary,
the continuum extrapolation is performed including lattice artefacts of $\O(a^2)$ and excluding the $N_t=6$ results.
The resulting continuum extrapolation is indicated by the shaded orange
band in Fig.~\ref{fig:mp-conti-extra}. We found this extrapolation to be more stable than the fit  to all available lattice spacings, including lattice
artefacts of $\O(a^4)$. Both extrapolations lead to similar results, see the 
comparison in Fig.~\ref{fig:mp-conti-extra}.

\begin{figure}[t]
 \centering
 \includegraphics[width=8cm]{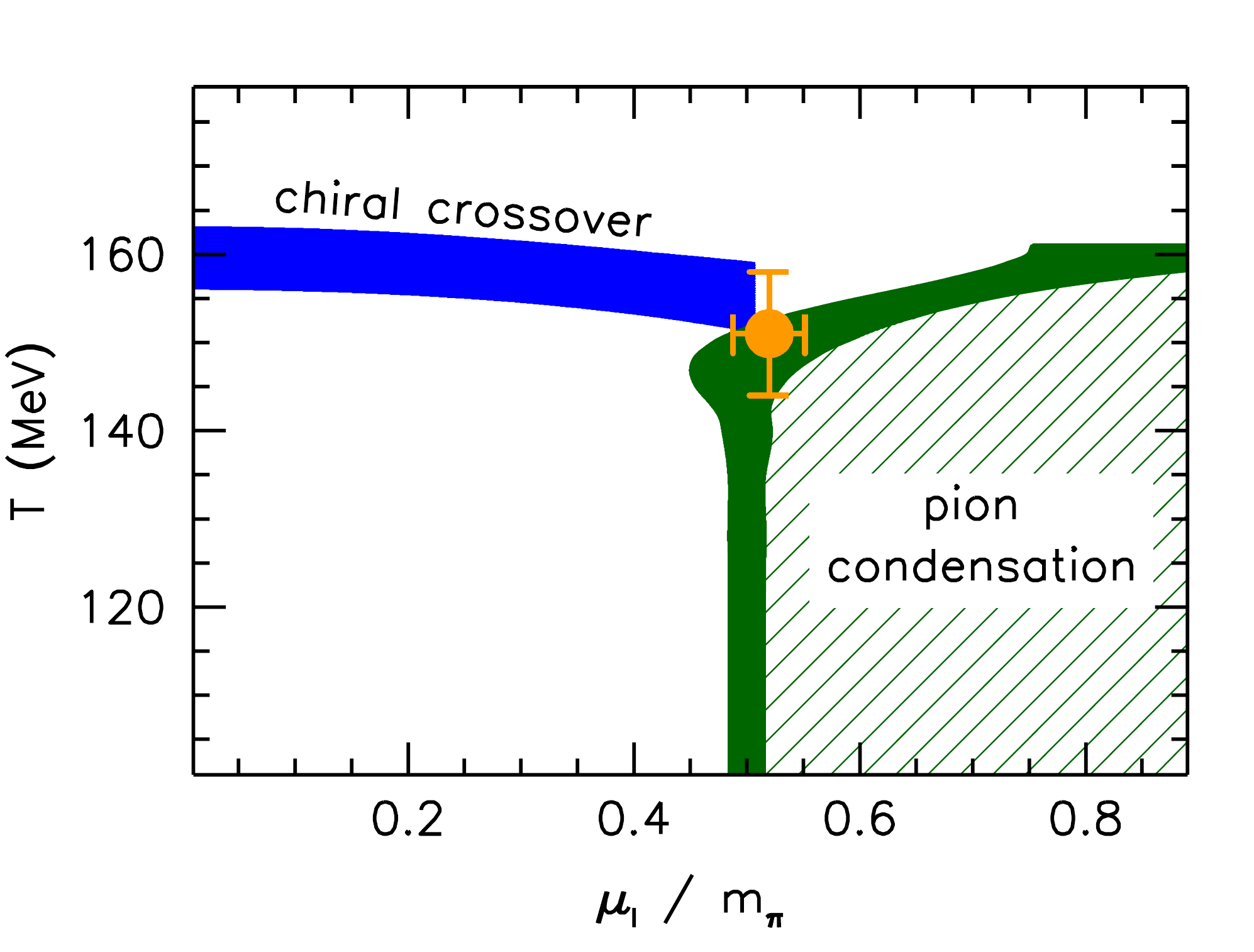}
 \caption{\label{fig:pd-final}
 The QCD phase diagram for nonzero isospin chemical potential in the continuum limit. 
 The blue band indicates the chiral crossover transition
 temperature $\Tc(\mu_I)$ and the green line is the boundary $\mu_{I,c}(T)$ of the pion condensation phase (the shaded green area). 
 The yellow point marks the triple point, beyond which 
 the two transitions are coincident (see text).
 }
\end{figure}

We are now in the position to draw the continuum phase diagram, which we display in Fig.~\ref{fig:pd-final}. The chiral crossover
transition starts from a temperature of $\Tc(0)=159(4)$~MeV, which is consistent with the crossover temperatures
from~\cite{Borsanyi:2010bp,*Bazavov:2011nk} within uncertainties. The results exhibit a small downward curvature
of the $\Tc(\mu_I)$ line. 
The pion condensation boundary remains at $\mu_{I,c}=m_\pi/2$ within our errors up to 
$T\approx140 \textmd{ MeV}$, beyond which it 
soon becomes very flat. For $T\gtrsim 160$~MeV, we do not observe pion condensation up to $\mu_I=120$~MeV.

\begin{figure}[t]
 \centering
 \includegraphics[width=8cm]{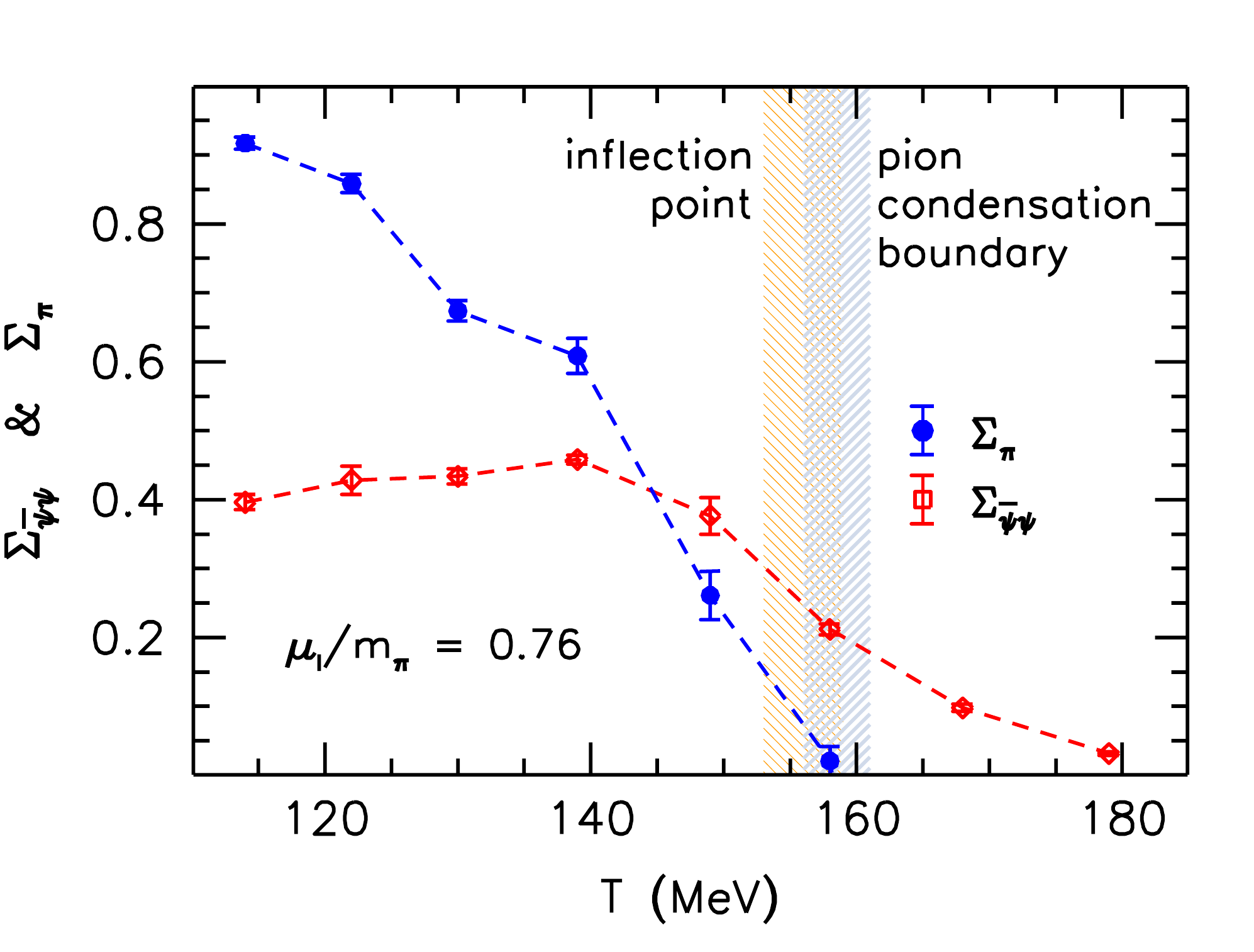}
 \caption{\label{fig:pcb-condens}
 Pion and quark condensates as functions of the temperature for $\mu_I=103$~MeV as measured on our $N_t=10$
 ensembles. The light blue area marks the pion condensation phase boundary and the orange area indicates the location of the
 inflection point of the condensate. The lines connecting the points are only
 included to guide the eye.
 }
\end{figure}

The two transition lines meet at the pseudo-triple point,
for which we obtain $\mut=70(5)$~MeV
in the continuum limit, indicated by the yellow point in Fig.~\ref{fig:pd-final}. 
The corresponding temperature is determined conservatively 
by taking into account the upper bound for
$\Tc(\mu_I=\mut)$ and the lower bound for the temperature 
where $\mu_{I,c}=\mut$. 
Defining the central value of $\Tt$ as the midpoint of this interval we obtain $\Tt=151(7)$~MeV.
From what we observe at
finite lattice spacings, we expect that chiral symmetry restoration and the pion condensation phase boundary coincide from
the pseudo-triple point on. To demonstrate this, in Fig.~\ref{fig:pcb-condens} we plot the pion condensate
together with the quark condensate for $\mu_I>\mut$. 
The figure indicates that pion condensation (defined by the point where $\Sigma_\pi=0$) occurs together with chiral symmetry
restoration (the inflection point of the condensate).
The initial rise of the chiral condensate in Fig.~\ref{fig:pcb-condens} is an interesting feature in
the pion condensation phase. A similar tendency has been observed in a study of the phase diagram of a related
two-color NJL model~\cite{Ratti:2004ra}.
We interpret it as a remnant of the relation between pion and chiral condensate
$\Sigma_{\bar\psi\psi}^2 + \Sigma_\pi^2=1$, discussed in Sec.~\ref{sec:obsrenorm}, which follows from
$\chi$PT to leading order~\cite{Son:2000xc}.

\subsection{Polyakov loop}

\begin{figure}[t]
 \centering
 \includegraphics[width=8cm]{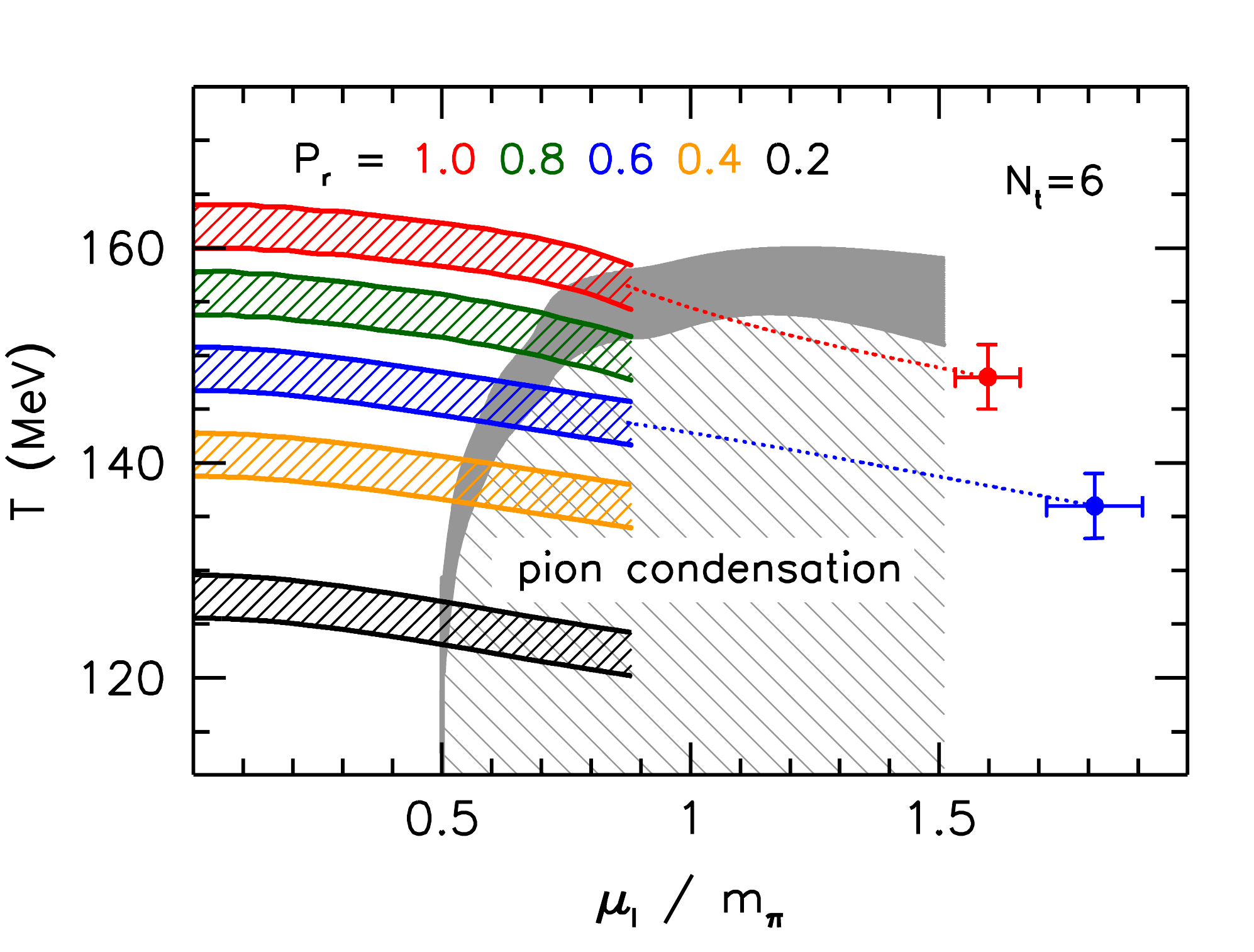}
 \caption{\label{fig:ploop-cont}
 Contour lines of constant renormalized Polyakov loop $P_r$ in the $\mu_I-T$ plane for the $N_t=6$ lattices.
 Here we extended our range of $\mu_I$-values and included two additional points beyond $\mu_I=120$~MeV obtained
 from an interpolation of $P_r$ for lines of constant temperature.
 }
\end{figure}

Next, we elaborate on the properties of the deconfinement transition in terms of the 
renormalized Polyakov loop $P_r$.
In contrast to the quark condensate, the Polyakov loop exhibits no pronounced inflection point. 
To capture how deconfinement depends on the isospin chemical potential, 
we consider the curves in the $\mu_I-T$ plane, where $P_r=\textmd{const.}$ is satisfied.
Considering our definition~(\ref{eq:pren}) for the renormalization, 
the contour with $P_r=1$ is a possible choice for the transition temperature. 
In addition, the distance between the various contour lines is related to
the slope of $P_r$ around the transition point.
We show the contours
of constant $P_r$ in the phase diagram for
$N_t=6$ in Fig.~\ref{fig:ploop-cont}. 
The contour lines are observed to be insensitive to the BEC phase transition, as they 
enter the pion condensation phase without any sign of critical behavior. 
In this plot we also included results obtained at high isospin chemical potentials, up to
$\mu_I\approx 200\textmd{ MeV}$ for the pion condensation boundary and up to 
$\mu_I\approx 240\textmd{ MeV}$ for a few Polyakov loop contours. 
After crossing the phase boundary, the contours
continue to decrease, in agreement with the sketch of the phase diagram in Fig.~\ref{fig:pd}. We postpone a more detailed
discussion of the deconfinement transition to a future publication with an extended range in $\mu_I$.

\subsection{Order of the BEC transition}

\begin{figure}[ht!]
 \centering
 \includegraphics[width=8cm]{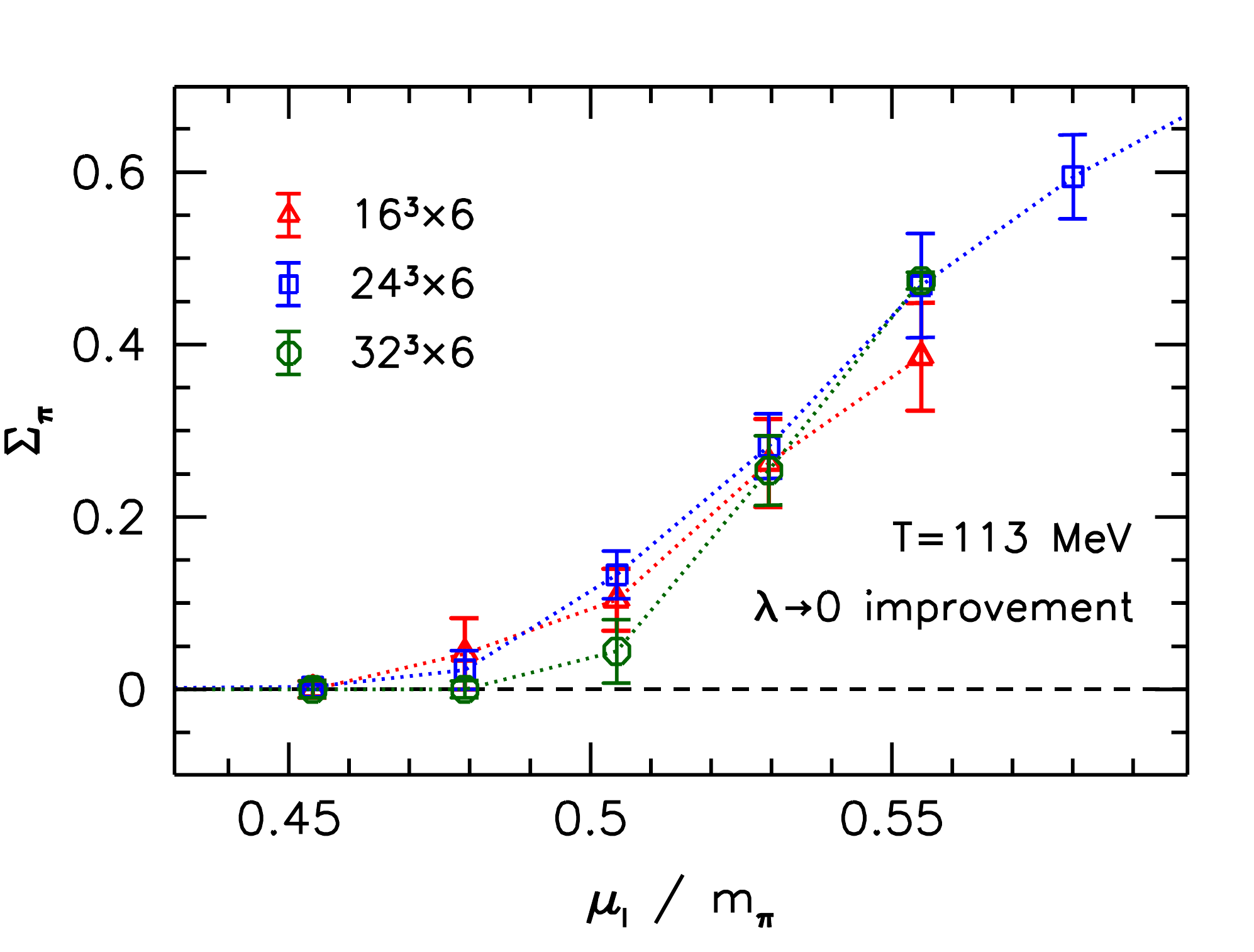}
 \includegraphics[width=8cm]{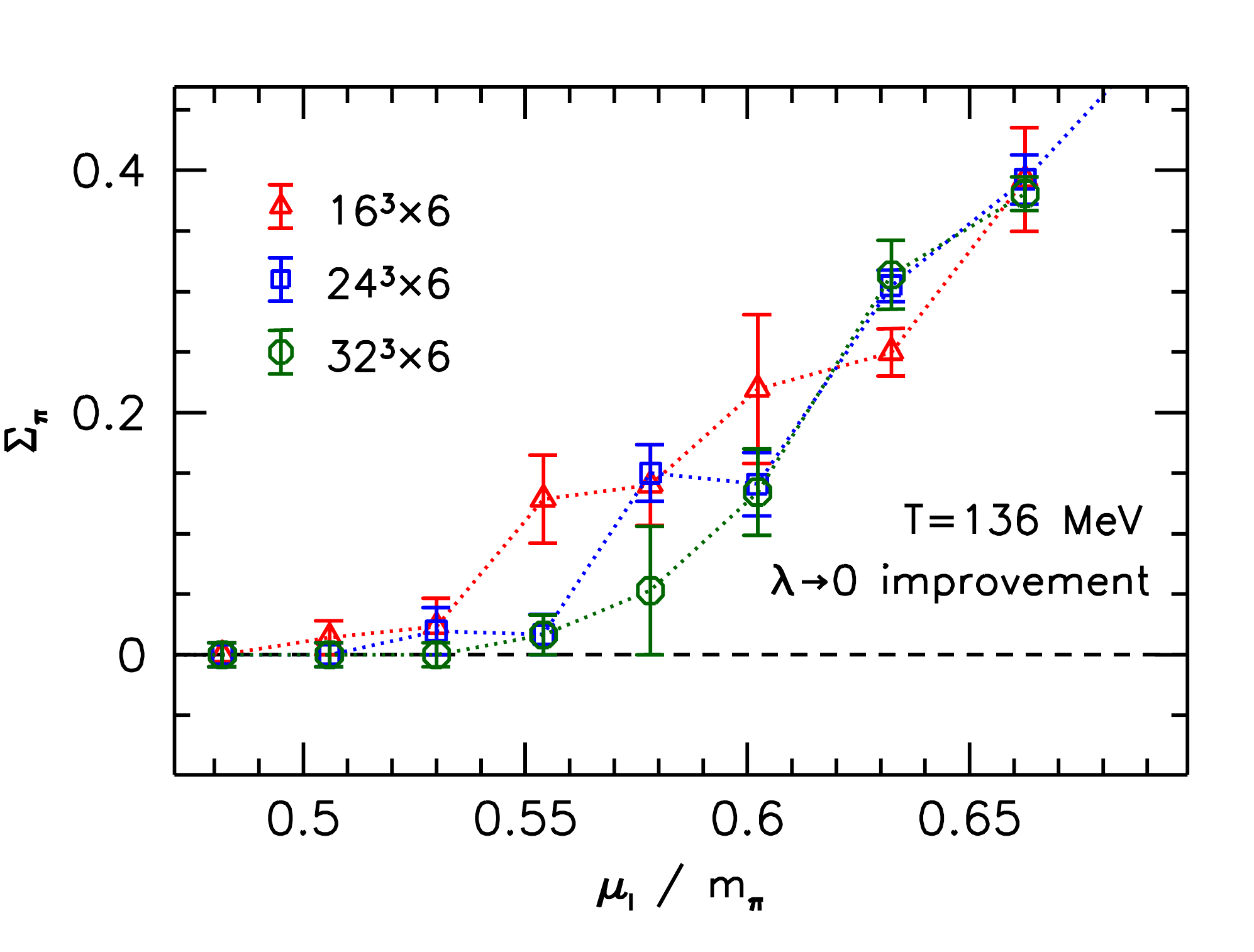}
 \caption{\label{fig:rho_voldep}
 Volume dependence of the ($\lambda\to0$ extrapolated) pion condensate $\Sigma_\pi$ around the critical isospin chemical potential
 for two temperatures well below (top panel) and somewhat below (bottom panel) the $\mu_I=0$ chiral
 transition temperature. The dotted lines merely serve to guide the eye. 
 }
\end{figure}

Finally, we investigate the nature of the transition between the vacuum and the pion condensed phase 
via a finite size scaling analysis. To this end 
three $N_t=6$ ensembles with $N_s=16$, $24$ and $32$ are employed. The results for $\Sigma_\pi$ 
around the critical isospin chemical potential $\mu_{I,c}$ are shown in Fig.~\ref{fig:rho_voldep}. 
Both at $T=113 \textmd{ MeV}$ and at $T=136\textmd{ MeV}$ a sharpening of
$\Sigma_{\pi}(\mu_I)$ is visible, 
confirming that a real phase transition takes place in the $V\to\infty$ limit. 
To be more quantitative, we proceed with the assumption that the transition is of second order.
In this case,
our results at $\lambda>0$ and $\mu_I\approx \mu_{I,c}$ 
must reflect the critical behavior that 
emerges as $\lambda\to0$ and $\mu_I\to\mu_{I,c}$ in 
the vicinity of the transition point. For the following analysis we use the data obtained on the $24^3\times 6$
ensemble at the lowest temperature $T=113\textmd{ MeV}$. 

\begin{figure}[b]
 \centering
 \includegraphics[width=8cm]{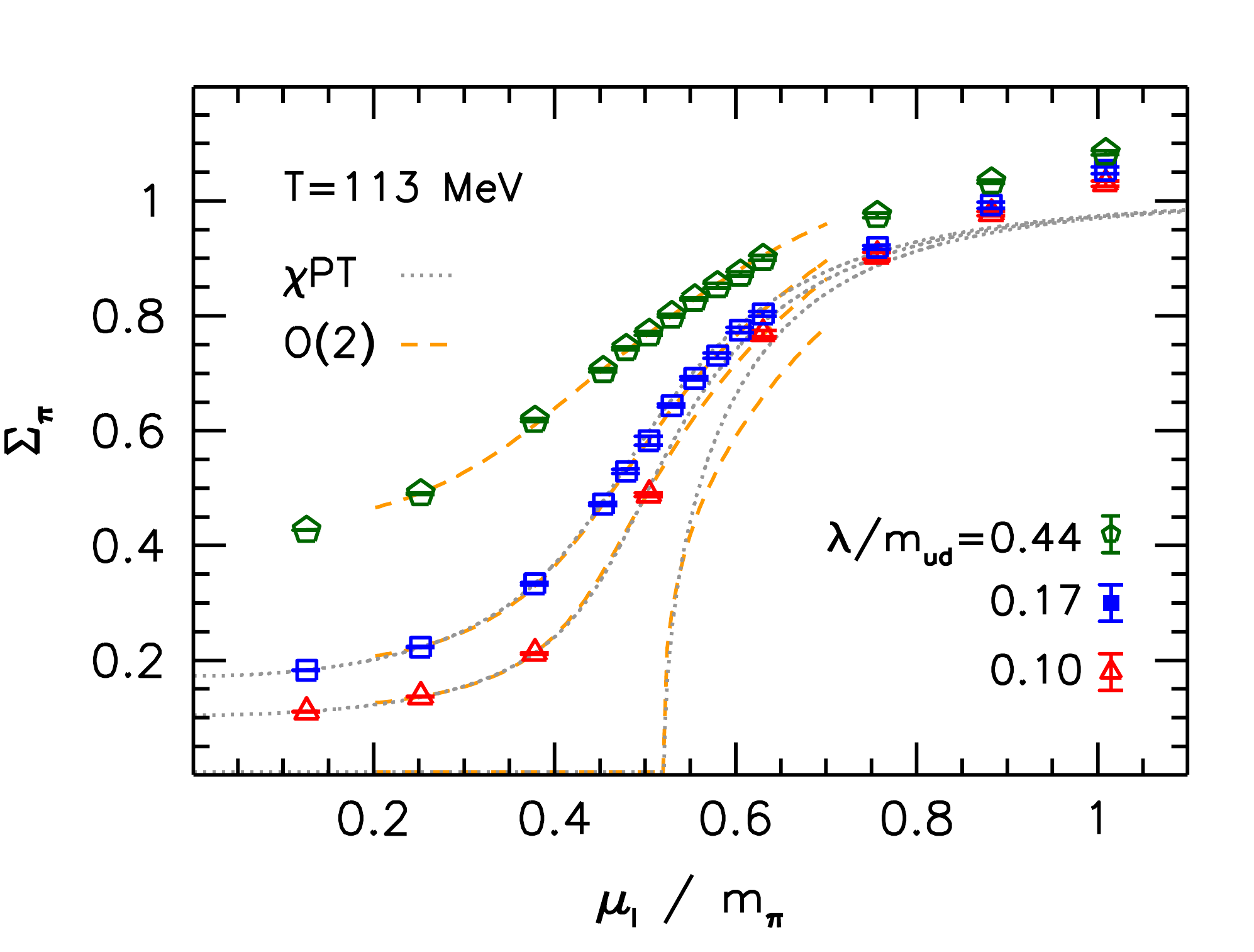}\quad
 \includegraphics[width=8cm]{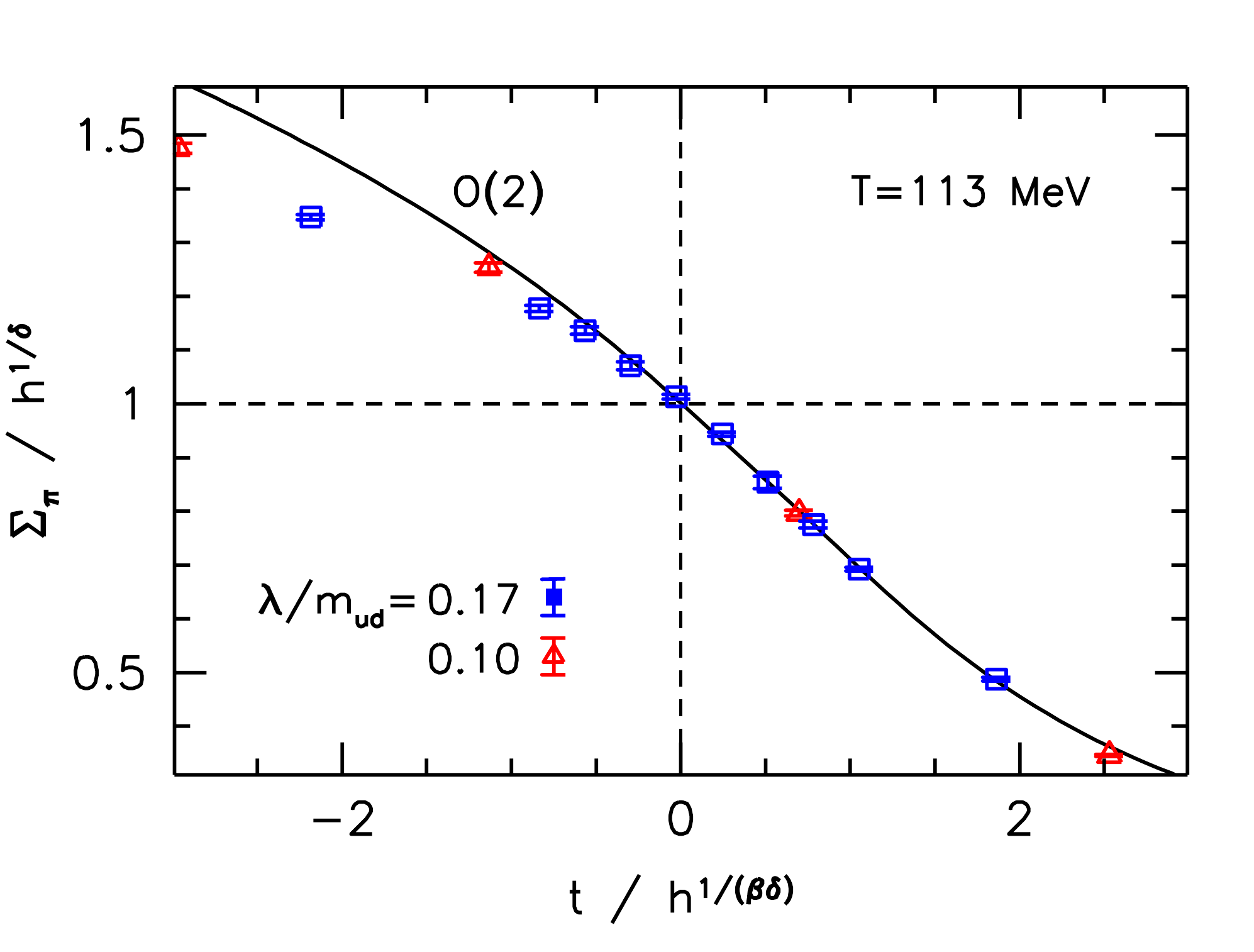}
 \caption{\label{fig:pbGpfits}
 Top panel: 
 comparison of our results for $\Sigma_\pi$ at nonzero $\lambda$ to $\chi$PT
 (dotted gray line)
 and to the critical behavior of the $\mathrm{O}(2)$ universality class including scaling 
 violations (dashed yellow line). 
 Bottom panel: collapse plot using the variables scaled according to the $\mathrm{O}(2)$ behavior,
 Eq.~(\protect\ref{eq:O2scaling}). 
 }
\end{figure}

First we compare our 
(unimproved) $\lambda>0$ results for $\Sigma_\pi$ to the prediction of $\chi$PT. 
This involves two free parameters: the critical isospin chemical potential and a normalization factor $G$, which corresponds to 
the magnitude of the renormalized quark condensate $\Sigma_{\bar\psi\psi}$ in the vacuum,
\be
\Sigma_\pi = G\sin\alpha\,,\quad
\sin\left( \alpha - \arctan\frac{\lambda}{\ml} \right) = \frac{2\mu_I^2}{\mu_{I,c}^2}\sin 2\alpha\,,
\ee
where the vacuum angle $\alpha$ is determined implicitly by the second equation (see Refs.~\cite{Splittorff:2002xn,Endrodi:2014lja}).
The data is very well described by the predicted behavior, see the top panel of 
Fig.~\ref{fig:pbGpfits}. 
Taking into account data points for $\mu_I/m_\pi<0.63$ and the two smallest 
pionic sources $\lambda/\ml<0.2$,  
we obtain a reasonable fit with $\chi^2/{\rm dof}\approx1.6$. 
For higher isospin chemical potentials $\chi$PT tends to 
underestimate $\Sigma_\pi$, as has already been noted in Ref.~\cite{Endrodi:2014lja}. 
The critical chemical potential is $\mu_{I,c}=71(2)\textmd{ MeV}$ and the normalization 
factor $G=1.02(1)$, both very close to the values expected in the zero-temperature and 
thermodynamic limits ($m_\pi/2$ and unity, respectively). 

We also consider the critical behavior of the $\mathrm{O}(2)$ universality class to fit 
the order parameter as a function of $\mu_I$ and $\lambda$. In particular, 
this dependence involves the universal scaling function $f_G$,
\be
\Sigma_\pi = h^{1/\delta} \cdot f_G\left(t/h^{1/(\beta\delta)}\right), \quad h=\frac{\lambda}{\lambda_0},
\quad t = \frac{\mu_{I,c}-\mu_I}{t_0}\,,
\label{eq:O2scaling}
\ee
with the critical exponents $\beta$ and $\delta$ and the 
free parameters $\mu_{I,c}$, $\lambda_0$, and $t_0$. 
For the construction of $f_G$, we follow Ref.~\cite{Ejiri:2009ac}. The result of
the fit, using the two lowest pionic sources and isospin chemical potentials $0.4<\mu_I/m_\pi<0.63$, is shown
in the bottom panel of 
Fig.~\ref{fig:pbGpfits}, demonstrating the collapse of the data on the universal curve. 
The critical chemical potential is found to be $\mu_{I,c}=74(2)\textmd{ MeV}$. However, 
since the fit quality is rather low ($\chi^2/{\rm dof}\approx 3$), we extended the 
fit function to include scaling violations in the spirit of Ref.~\cite{Ejiri:2009ac}. We found it necessary to work with
\be
\Sigma_\pi = h^{1/\delta} \cdot f_G\left(t/h^{1/(\beta\delta)}\right) + a_1 t h + b_1 h + b_3 h^3\,,
\ee
which allowed us to achieve reasonable fit qualities ($\chi^2/{\rm dof}\approx 1.5$) 
for broader fit intervals $\lambda/\ml<0.5$ and $0.2<\mu_I/m_\pi<0.7$. 
The resulting fit is included in the top panel of Fig.~\ref{fig:pbGpfits}. 
We find $\mu_{I,c}=70(2) \textmd{ MeV}$, again lying very close to the expected 
value of $m_\pi/2$. 

Summarizing, our results strongly indicate that the transition into the pion condensed 
phase is of second order
in the infinite volume limit. In addition, the behavior of the order parameter at nonzero values of the 
explicit breaking parameter $\lambda$ is consistent with the critical scaling in the $\mathrm{O}(2)$ 
universality class, as expected due to the spontaneous symmetry breaking pattern discussed in Sec.~\ref{sec:sb}.

\section{Summary}

In this paper we have presented the first study of the phase diagram of QCD at nonzero isospin chemical
potentials in the continuum limit with dynamical $u$, $d$ and $s$ quarks at physical quark masses.
Our main result is the continuum extrapolated phase diagram, as shown in Fig.~\ref{fig:pd-final}.
The key features of the phase diagram
are the chiral crossover transition, the pion condensation boundary and their meeting point, 
the pseudo-triple point. We observe
that the critical chemical potential for pion condensation
remains constant up to $T\approx140$~MeV,
followed by a drastic change in behavior and a saturation at around
$T\approx160$~MeV. Above this temperature, we no longer observe
pion condensation, at least up to $\mu_I=120$~MeV. 
This finding might be particularly relevant for the orbifold equivalence in the large $N_c$ limit, 
as the pion condensation region is smaller than previously expected. 

The chiral crossover
temperature decreases slightly until it reaches the vicinity of the pion condensation phase boundary. The two
transition lines meet at $\mut=70(5)$~MeV and are found to be on top of each other within uncertainties
for higher isospin chemical potentials. 
Our scaling analysis of the pion condensation boundary shows consistency with
a second order phase transition in the $\mathrm{O}(2)$ universality class. 

Using our $N_t=6$ lattice ensembles, we find first indications that the deconfinement
transition temperature -- defined in terms of the renormalized Polyakov loop --
decreases and smoothly penetrates into the pion condensation phase. This behavior is 
depicted schematically in Fig.~\ref{fig:pd}. If this tendency persists in the continuum limit
and for larger values of $\mu_I$, the scenario where the deconfinement transition connects continuously to the
BEC-BCS crossover will be favored. Our results for the phase diagram 
can be compared directly to effective theories and models of QCD 
and serve as a highly nontrivial check of 
the validity of these approaches.

The main technical novelty of this study is the development of an improvement program for the extrapolations in the
infrared regulator $\lambda$, discussed in detail in Secs.~\ref{sec:imprl} and~\ref{sec:results}. Without the use of
these improved $\lambda$ extrapolations the reliable extraction of the phase diagram would have hardly been
possible. We would like to emphasize that a similar infrared regulator will likely be necessary to enable simulations
at finite baryon chemical potential -- granted that the sign problem has been solved. In this case, a generalization of
the methods presented here might be helpful to give insight to the investigations at finite $\mu_B$.

\begin{acknowledgments}
This research was funded by the DFG (Emmy Noether Programme EN 1064/2-1 and
SFB/TRR 55). The simulations were performed on the GPU cluster 
of the Institute for Theoretical Physics at the University of Regensburg 
and on the FUCHS and LOEWE clusters at the Center for Scientific Computing of the
Goethe University of Frankfurt.
The authors thank Gunnar Bali, Falk Bruckmann, Tom DeGrand, 
Simon Hands, S\'andor Katz, Andreas Sch{\"a}fer, Lorenz von Smekal, K\'alm\'an Szab\'o
and Tilo Wettig
for illuminating discussions. 
\end{acknowledgments}

\end{document}